\documentclass[10pt]{llncs} 
\usepackage{a4wide}
\usepackage{cite}
\usepackage{amsmath, amsxtra, amsfonts, amssymb}
\usepackage{eucal, url}
\usepackage{cite}
\newcommand{\imp}{\,{{:}{-}}\,}
\usepackage{xspace, varioref}

\DeclareMathOperator{\Id}{Id}

\newcommand{\manuel}[1]{\marginpar{\footnotesize{\emph{#1}}}}

\DeclareMathOperator{\Csp}{CSP}

\newcommand{\dlg}{{{:}{-}}}
\newcommand{\cproblem}[3]{
\vspace{.2cm}
\noindent {\bf #1} \\
INSTANCE: #2 \\
QUESTION: #3 \\}

\newcommand{\dom}{\text{\emph{dom}}}

\newcommand{\Fresse}{Fra\"{i}ss\'{e}}

\newcommand{\Qu}{\Gamma}

\newcommand{\tc}{{\it tc}}

\newcommand{\edgeS}{{\it edge}\xspace}

\newcommand{\false}{\ensuremath{{\it false}}\xspace}

\newcommand{\ignore}[1]{}

\begin{document}

\title{Datalog and Constraint Satisfaction \\ 
with Infinite Templates}

\author{Manuel Bodirsky\inst{1}\thanks{The first author has received funding from the European Research Council under the European Community's Seventh Framework Programme (FP7/2007-2013 Grant Agreement no. 257039)} \and V\'ictor Dalmau\inst{2}\thanks{Supported by the MICINN through grant TIN2010-20967-C04-02}}
\institute{CNRS/LIX, \'Ecole Polytechnique, \email{bodirsky@lix.polytechnique.fr}
\and Universitat Pompeu Fabra, \email{victor.dalmau@upf.edu}}

\maketitle


\begin{abstract}
On finite structures, there is a well-known connection 
between the expressive power of Datalog, finite variable logics, 
the existential pebble game,
and bounded hypertree duality.
We study this connection for infinite structures.
This has applications for constraint satisfaction with
infinite templates. 
If the template $\Gamma$ is $\omega$-categorical,
we present various equivalent characterizations of those 
$\Gamma$ such that the constraint satisfaction problem (CSP) for $\Gamma$ can be solved
by a Datalog program.
We also show that $\Csp(\Gamma)$
can be solved in polynomial time for arbitrary $\omega$-categorical structures 
$\Gamma$ if the input is restricted
to instances of bounded treewidth.
Finally, we characterize those $\omega$-categorical templates whose CSP has Datalog width $1$, 
and those whose CSP has strict Datalog width $k$.
\end{abstract} 

An extended abstract of this paper 
appeared in the proceedings of the 23rd International Symposium on Theoretical Aspects of Computer Science (STACS'06)~\cite{BodDal}. 


\section{Introduction}
\label{sect:intro}
In a constraint satisfaction problem we are given a set
of variables and a set of constraints on these variables, and want
to find an assignment of values from some domain $D$ 
to the variables such that all
the constraints are satisfied. The computational
complexity of a constraint satisfaction problem depends
on the type of constraints that can be used in the instances
of the problem. 
For \emph{finite} domains $D$, the complexity of the constraint
satisfaction problem 
attracted considerable attention in recent 
years; 
we refer to a recent collection of survey papers for a 
more complete account~\cite{CSPSurveys}.

Constraint satisfaction problems where the domain $D$ is infinite
have been studied in Artificial Intelligence and the
theory of binary relation algebras~\cite{LadkinMaddux,Duentsch},
with applications for instance in temporal and spatial reasoning.
Well-known examples of such binary relation algebras are
the \emph{point algebra}, the \emph{containment algebra}, 
\emph{Allen's interval algebra}, and the \emph{left linear point algebra};
see~\cite{HirschCristiani,HirschAlgebraicLogic,LadkinMaddux,Duentsch} and the references therein.

Constraint satisfaction problems can be modeled as homomorphism 
problems~\cite{FederVardi}. 
For detailed formal definitions of relational structures and
homomorphisms, see Section~\ref{sect:defs}, and for
the connection to network satisfaction problems for relation algebras, see Section~\ref{sect:relation-algebra}. 
Let $\Gamma$ be a (finite or infinite) structure 
with a finite relational signature $\tau$. 
Then the \emph{constraint satisfaction problem (CSP)} 
for $\Gamma$ is the following computational problem. 

\cproblem{CSP($\Gamma$)}
{A finite $\tau$-structure $A$.}
{Is there a homomorphism from $A$ to $\Gamma$?}

The structure $\Gamma$ is called the \emph{template} of the constraint
satisfaction problem $\Csp(\Gamma)$. 
For example, if the template is the dense linear order of the rational
numbers $(\mathbb{Q}, <)$, then it is easy to see that
$\Csp(\Gamma)$ is the well-known problem of digraph-acyclicity.

Many constraint satisfaction problems 
in Artificial Intelligence 
can be formulated with \emph{$\omega$-categorical} templates. 
The concept of $\omega$-categoricity is of central importance
in model theory and will be introduced in Section~\ref{sect:omegacat}; in the context of the network satisfaction problems for relation algebras, the relevance of $\omega$-categoricity has already been recognized in~\cite{HirschAlgebraicLogic}. 
An important class of examples for $\omega$-categorical structures
are the so-called \Fresse-limits of amalgamation classes with finite relational signature~\cite{HodgesLong}.
It is well-known that all the CSPs 
for the binary relation algebras (and their fragments) mentioned 
above, and many other problems in temporal and spatial reasoning can be formulated with $\omega$-categorical structures.

For $\omega$-categorical templates we can apply the so-called 
\emph{algebraic approach 
to constraint satisfaction}~\cite{JeavonsClosure,JBK,CSPSurvey}
to analyze the computational complexity of the corresponding CSPs.
This approach was originally developed for constraint
satisfaction with finite templates, but several fundamental facts of the universal-algebraic approach also hold for $\omega$-categorical templates~\cite{BodirskyNesetrilJLC,Cores-journal}. The universal-algebraic approach has
been used to obtain complete complexity classifications for
large classes of $\omega$-categorical templates~\cite{tcsps-journal,BodPin-Schaefer}.


\paragraph{Datalog.}
Datalog is an important algorithmic tool to study the complexity of CSPs. It can be viewed as 
the language of logic programs without function
symbols, see e.g.~\cite{KolaitisVardi, EbbinghausFlum}.
For constraint satisfaction with finite domains, 
Datalog was first investigated systematically 
by Feder and Vardi~\cite{FederVardi}.
Also for CSPs with infinite domains, Datalog programs 
play an important role (even though this is usually not made
explicit in the literature),  because
one of the most studied algorithms in infinite-domain constraint satisfaction, the 
\emph{path consistency algorithm}, 
and many of its variants
can be formulated by Datalog programs (see Section~\ref{sect:relation-algebra}).

Fix a set of relation symbols $\sigma$. 
A Datalog program consists of a finite set of 
\emph{rules}, traditionally written 
in the form $$\phi_0 \imp \phi_1, \dots, \phi_r$$
where $\phi_0,\phi_1,\dots,\phi_r$ are
\emph{atomic $\sigma$-formulas}, that is, formulas
of the form $R(x_1,\dots,x_n)$ for $R \in \sigma$ and
variables $x_1,\dots,x_n$. 
In such a rule $\phi_0$ is called the \emph{head}
and $\phi_1,\dots,\phi_r$ the \emph{body} of the rule.
The relation symbols that never appear 
in rule heads are called the \emph{input relation symbols},
or \emph{EDBs} (this term comes from database theory, and stands for \emph{extensional database}). 
The other relation symbols that appear in the Datalog program are called \emph{IDBs} 
(short for \emph{intentional database}).

Before we give formal definitions of the semantics of a Datalog program in Section~\ref{sect:defs},
we show an instructive example. 
\begin{align*}
  {\it tc}(x,y) \; & {{:}{-}} \; {\it edge}(x,y) \\
  {\it tc}(x,y) \; & {{:}-} \; {\it tc}(x,u), {\it tc}(u,y) \\
  \false \; & {{:}-} \; {\it tc}(x,x)
\end{align*}
Here, the binary relation \edgeS is the only input relation symbol, \tc~is a binary IDB, and \false is a $0$-ary IDB. 
Informally, the Datalog program computes with the help of the relation \tc~the transitive closure 
of the edges in the input relation, and 
derives \false if and only if the input (which can be seen as a digraph
defined on the variables) contains a directed cycle.
Hence, the program above derives \false on a given directed graph if and only if
the directed graph does \emph{not} homomorphically map to $({\mathbb Q};<)$.
In general, we say that a CSP is \emph{solved} by a Datalog program 
if the distinguished $0$-ary predicate \false is derived on an instance of the CSP if and only if the instance has no solution. This will be made precise in Section~\ref{sect:defs}.

An important measure for the complexity of a Datalog
program is the maximal number $k$ of variables per rule (see e.g.~\cite{GroheThesis}). On structures of size $n$,
such a Datalog program
can be evaluated in time $O(n^{k+1})$. 
(Hence, a fixed Datalog program can be evaluated in time polynomial in $n$.)
In this work, we are interested in capturing a finer distinction,
and study the expressive power of Datalog depending 
both on the maximal number $k$ of variables per rule and on the maximal number $l$ of variables \emph{in the head} of the rules. Such Datalog programs are said to have
\emph{width $(l,k)$}. The Datalog program shown above,
for instance, has width $(2,3)$.
The double parameterization 
is less common, but more general, and has already been considered 
in the literature on constraint satisfaction and Datalog~\cite{FederVardi}.

For finite templates $\Gamma$, 
it has been shown that there is a tight connection
between the expressive power of Datalog, the so-called existential pebble game,  finite variable logics, and bounded hypertree duality;
these concepts will be introduced in Section~\ref{sect:defs}
and the mentioned connection will be formally stated in Section~\ref{sect:negative}.
The connection shows that the following are equivalent:
\begin{itemize}
\item there is a Datalog program of width $(l,k)$
that solves $\Csp(\Gamma)$;
\item for all instances $A$ of $\Csp(\Gamma)$, 
if 
Duplicator has a winning strategy strategy for the existential $(l,k)$-pebble game on $A$ and $\Gamma$,
then $A$ homomorphically maps to $\Gamma$;
\item for all instances $A$ of $\Csp(\Gamma)$, 
 if all sentences in the infinitary $l$-bounded existential positive $k$-variable logic $L^{l,k}_{\infty \omega}$
that hold in $A$ also hold in $\Gamma$,
then $A$
homomorphically
maps to $\Gamma$;
\item there is a set $\mathcal N$ of finite structures of 
treewidth at most $(l,k)$ such that every finite $\tau$-structure
$A$ is homomorphic to $\Gamma$ is and only if no structure in $\cal N$ is homomorphic to $A$.
\end{itemize}
The four characterizations provide links to different research areas -- database theory, constraint satisfaction complexity, finite model theory, combinatorics -- and the possibility
to change between the various perspectives on width-bounded Datalog has had a profound
impact on research in those areas.

We mention that recently an (effective) universal-algebraic condition has been found that characterizes which finite structures $\Gamma$ have a CSP that can be solved with Datalog~\cite{BoundedWidth,Bulatov-BoundedWidth}. A corresponding characterization of those $\omega$-categorical templates that can be solved by Datalog remains open. 

\paragraph{Results.}
We study the connection 
between the expressive power of Datalog, 
finite variable logics, the existential pebble game,
and bounded hypertree duality for \emph{infinite structures $\Gamma$}. We show that the result for finite structures $\Gamma$ mentioned above
fails for general infinite structures (Section~\ref{sect:negative}), 
but holds true if $\Gamma$ is $\omega$-categorical (Section~\ref{sect:datalog-omegacat}).
An important tool to characterize the expressive 
power of Datalog for constraint satisfaction 
is the notion of \emph{canonical Datalog programs}.
This concept was introduced by Feder and Vardi for finite templates;
we present a generalization to $\omega$-categorical templates.
We prove that a CSP with an $\omega$-categorical
template can be solved
by an $(l,k)$-Datalog program if and only if the 
canonical $(l,k)$-Datalog program for $\Gamma$ solves the problem 
(Section~\ref{sect:canonical}).

An important consequence of our result is that for $\omega$-categorical $\Gamma$, the problem
$\Csp(\Gamma)$
can be solved in polynomial time if the input is restricted
to a class of instances of bounded treewidth (in fact, it suffices that the
\emph{cores} of the instances have bounded treewidth).

We also investigate which CSPs
can be solved with a Datalog program (and are thus polynomial-time tractable) 
when  no restriction is imposed on the input instances 
(Section~\ref{sect:bounded}).
In particular, we prove a characterization of CSPs
with $\omega$-categorical
templates $\Gamma$ that can be solved by a Datalog program of width $(1,k)$. 
In fact, every problem that
is closed under disjoint unions and can be solved by a Datalog program of width $(1,k)$ for some $k$
can be formulated as a constraint
satisfaction problem with an $\omega$-categorical template.
More generally, one can find $\omega$-categorical templates
for problems that are closed under disjoint unions and 
can be described in the logic called \emph{monotone monadic SNP} introduced by Feder and Vardi~\cite{FederVardi}) 
(Section~\ref{sect:datalogascsp}); to show this, we apply a model-theoretic result of Cherlin, Shelah, and Shi~\cite{CherlinShelahShi}.

A special case of width $(1,k)$-Datalog programs are
problems that can be decided by establishing \emph{arc-consistency} 
(sometimes also called \emph{hyperarc-consistency}),
which is a well-known and intensively studied 
technique in artificial intelligence. 
We show that if a constraint satisfaction problem with an $\omega$-categorical
template can be decided by establishing arc-consistency, 
then it can also be formulated as a constraint satisfaction problem
with a finite template (Section~\ref{sect:bounded}).

Finally, we characterize \emph{strict width $l$}, a notion that was again 
introduced for finite templates and for $l\geq 2$ in~\cite{FederVardi}.
Intuitively, $\Csp(\Gamma)$ has strict width $l$, for $l \geq 2$,
if there is a Datalog program of width $(l,k)$, for some
$k \geq l+1$, that computes on a given
instance $A$ of $\Csp(\Gamma)$ \emph{`all the $l$-ary facts that are implied by $A$'}, that is, it makes all semantically
entailed $l$-ary constraints syntactically present. Obviously,
this needs a careful formal definition, which we present
in Section~\ref{sect:swk}.
Jeavons et al.~\cite{CCC} 
say that in this case \emph{establishing
strong $l$-consistency ensures global consistency}. 
For finite templates, strict width $l$ can be characterized by
an algebraic closure condition~\cite{FederVardi,CCC}. 
In Section~\ref{sect:swk} we generalize 
this result to $\omega$-categorical templates $\Gamma$ with a finite signature, and show
that $\Csp(\Gamma)$ has strict width $l$ if and only if for 
every finite subset $A$ of the domain of
$\Gamma$ there is an $(l+1)$-ary polymorphism of $\Gamma$
that is a \emph{near-unanimity
operation on $A$}, i.e., it satisfies the identity 
$f(x, \dots, x, y, x, \dots, x) = x$ for all $x,y \in A$.

\section{Definitions and basic facts}
\label{sect:defs}
A \emph{relational signature} $\tau$ is a (here always at most countable) 
set of \emph{relation symbols} $R_i$ (also called \emph{predicates}), each associated with an \emph{arity} 
$k_i \in {\mathbb N}$. A {\em (relational) structure} $\Qu$ \emph{over relational signature} $\tau$ (also called $\tau$-\emph{structure}) is a set $D_\Qu$ (the {\em domain}) together with a relation $R_i \subseteq D_\Qu^{k_i}$ for each relation symbol of arity $k_i$. If necessary, we write $R^\Gamma$ to indicate that we are talking about the relation $R$ belonging to the structure $\Gamma$. For simplicity, we otherwise denote both a relation symbol and its corresponding relation with the same symbol. 
For a $\tau$-structure $\Qu$ and $R\in \tau$ it will also be convenient to say that $R(u_1, \dots, u_k)$ \emph{holds in} $\Qu$ iff $(u_1, \dots, u_k) \in R$. We sometimes use the shortened notation $\overline{x}$ for a vector $x_1, \dots, x_n$ of any length.
If we add relations to a given $\tau$-structure $\Gamma$,
then the resulting structure $\Gamma'$ with a larger signature 
$\tau' \supset \tau$ is called a \emph{$\tau'$-expansion} of $\Gamma$, 
and $\Gamma$ is called a \emph{$\tau$-reduct} of $\Gamma'$. 


The \emph{union} of two $\tau$-structures $\Gamma$ and $\Gamma'$ with disjoint 
domains
is a $\tau$-structure $\Delta$ that is defined on the union of the
domains of $\Gamma$ and $\Gamma'$.
We have $R^\Delta := R^\Gamma \cup R^{\Gamma'}$ for every $R \in \tau$. 
When the domain of $\Gamma$ and $\Gamma'$ is not disjoint, a \emph{disjoint union} $\Delta$ of $\Gamma$ and $\Gamma'$ is the union of $\Gamma$ with a copy of $\Gamma'$ 
whose domain is distinct from the domain of $\Gamma$.
Since we usually consider structures up to isomorphism, we also call the structure $\Delta$
\emph{the} disjoint union of $\Gamma$ and $\Gamma'$. 

A $\tau$-structure is called \emph{connected} iff it is not the
disjoint union of two $\tau$-structures with a non-empty domain.
The \emph{Gaifman graph} (sometimes also called the \emph{shadow}) of
a relational structure $A$ is a graph on the vertex set $v_1,\dots,v_n$
where two distinct vertices $v_k$ and $v_l$ are adjacent if 
there is a relation in $A$ that is imposed on both $v_k$ and $v_l$, i.e.,
there is a relation $R$ such that 
$A$ satisfies $R(v_{i_1},\dots,v_{i_j})$
and $k,l \in \{i_1,\dots,i_j\}$. It is clear that a structure is connected if and only if its Gaifman graph is connected.  

\subsection{Homomorphisms}
Let $\Gamma$ and $\Delta$ be $\tau$-structures. A \emph{homomorphism} from $\Gamma$ to $\Delta$ is a function $f$ from $D_{\Gamma}$ to $D_{\Delta}$ such that for each $n$-ary relation symbol $R$ in $\tau$ 
and each $n$-tuple ${\overline a}=(a_1,\dots,a_n)$, if ${\overline a} \in R^\Gamma$, then $(f(a_1), \dots, f(a_n)) \in R^{\Delta}$. In this case we say that the map $f$ \emph{preserves} the relation $R$.
Two structures $\Gamma$ and $\Delta$ 
are called \emph{homomorphically equivalent} if there is a homomorphism
from $\Gamma$ to $\Delta$ and a homomorphism from $\Delta$ to $\Gamma$.

A \emph{strong homomorphism} $f$ satisfies the stronger condition that for each $n$-ary relation symbol in $\tau$ and each $n$-tuple $\overline a$
we have that ${\overline a} \in R^\Gamma$ if and only if $(f(a_1), \dots, f(a_n)) \in R^{\Delta}$.
An \emph{embedding} of a structure $\Gamma$ into a structure $\Delta$ is an injective strong homomorphism. An \emph{isomorphism} is a surjective embedding. Isomorphisms from $\Gamma$ to $\Gamma$ are called \emph{automorphisms}.

If $\Gamma$ and $\Delta$ are structures of the same signature, 
with $D_\Gamma \subseteq D_{\Delta}$, and the inclusion map is an embedding, 
then we say that $\Delta$ is an \emph{extension} of $\Gamma$, 
and that $\Gamma$ a \emph{restriction} of $\Delta$. 

A partial mapping $h$ from a relational structure $A$ to a relational structure
$B$ is called a \emph{partial homomorphism} (from $A$ to $B$)
if $h$ is a homomorphism from
a restriction $A'$ of $A$ to $B$
and $A'$ is the domain of $h$.
As usual, the \emph{restriction} of a function $h$ to a
subset $S$ of its range 
is the mapping $h'$ with range $S$ where $h'(x)=h(x)$ for all $x \in S$; $h$ is called an \emph{extension} of $h'$.


\subsection{First-order logic}
First-order formulas $\varphi$ over the signature $\tau$ (or, in short, $\tau$-formulas) are inductively defined using the logical symbols of universal and existential quantification, disjunction, conjunction, negation, equality, bracketing, variable symbols and the symbols from $\tau$. The semantics of a first-order formula over some $\tau$-structure is defined in the usual Tarskian style. A $\tau$-formula without free variables is called a $\tau$-sentence. We write $\Gamma \models \varphi$ iff the $\tau$-structure $\Gamma$ is a model for the $\tau$-sentence $\varphi$; this notation is lifted to sets of sentences in the usual way. A good introduction to logic and model theory is~\cite{HodgesLong}.

We can use first-order formulas over the signature $\tau$ to
define relations over a given $\tau$-structure $\Gamma$: 
for a formula 
$\varphi(x_1,\dots,x_k)$ where $x_1,\dots,x_k$ are the
free variables of $\varphi$ the corresponding relation $R$ is
the set of all $k$-tuples $(t_1,\dots,t_k) \in D_\Gamma^k$ such that $\varphi(t_1,\dots,t_k)$ is true in $\Qu$.

A first-order formula $\varphi$
is said to be \emph{primitive positive} 
(we say $\varphi$ is a \emph{pp-formula}, for short) iff it is of the form 
$\exists \overline{x} \; (\varphi_1({\overline x}) \wedge \dots \wedge \varphi_k({\overline x}))$ 
where $\varphi_1, \dots, \varphi_k$ are atomic formulas (which might be
equality relations of the form $x=y$). 

\subsection{Canonical queries}\label{ssect:queries}
A basic concept to link structure homomorphisms and logic is the
\emph{canonical conjunctive query $\phi^A$} of a finite relational structure $A$,
which is a first-order formula of the form $\exists v_1,\dots,v_n \; (\psi_1 \wedge \dots \wedge \psi_m)$, where $v_1,\dots,v_n$ are the vertices of $A$,
and $\{\psi_1,\dots,\psi_m\}$ is the set of atomic formulas of the
form $R(v_{i_1},\dots,v_{i_j})$ that hold in $A$.

It is a fundamental property of the canonical query $\phi^A$ that 
$\phi^A$ holds in a structure $\Gamma$ if and only if there
is a homomorphism from $A$ to $\Gamma$~\cite{ChandraMerlin}.

\subsection{Finite variable logics}
The class of sentences that have at most $k$ variables and are obtained from atomic formulas using infinitary
conjunction, infinitary disjunction, and existential quantification is denoted by 
$\exists {\cal L}^k_{\infty \, \omega}$.
The class $\bigcup_{k\geq 0}\exists {\cal L}^k_{\infty \, \omega}$
is denoted by $\exists {\cal L}^\omega_{\infty \, \omega}$.

We want to bring another parameter $l$ into the picture,
and define the following refinement of $\exists {\cal L}^k_{\infty\, \omega}$. A conjunction 
$\bigwedge \Psi$ is called \emph{$l$-bounded}
if $\Psi$ is a collection of $\exists {\cal L}^\omega_{\infty \, \omega}$ formulas $\psi$ 
that are quantifier-free or have at most $l$ free variables.
Similarly, an disjunction 
$\bigvee \Psi$ is called \emph{$l$-bounded}
if $\Psi$ is a collection 
of $\exists {\cal L}^\omega_{\infty \, \omega}$ formulas $\psi$ 
that are quantifier-free or have at most $l$ free variables.
The set of $\exists {\cal L}^{l,k}_{\infty\, \omega}$
formulas is defined as the restriction of $\exists {\cal L}^k_{\infty \, \omega}$ obtained by only allowing
infinitary $l$-bounded conjunction and $l$-bounded disjunction instead of full infinitary conjunction and disjunction. Note that 
$\bigcup_{0\leq l<k} \exists {\cal L}^{l,k}_{\infty \, \omega}$ equals
$\exists {\cal L}^k_{\infty \, \omega}$. The logic $\exists {\cal L}^k_{\infty \, \omega}$ was introduced (under a different name) by Kolaitis 
and Vardi as an existential negation-free 
variant of well-studied infinitary logics to study the expressive power of Datalog~\cite{KolaitisVardiDatalog}; in subsequent work, they
used the name $\exists {\cal L}^k_{\infty \, \omega}$ to denote this 
logic~\cite{KolaitisVardi} and we follow this convention.

We denote by $L^{l,k}$ the logic
$\exists {\cal L}^{l,k}_{\infty \, \omega}$ without disjunctions
and with just \emph{finite} $l$-bounded conjunctions.
In other words, a formula in $L^{l,k}$ is composed out of
existential quantification and finitary $l$-bounded conjunction, and
uses only $k$ distinct variables. We also
call the logic $L^{l,k}$ \emph{infinitary $l$-bounded existential positive $k$-variable logic}.
The language $L^k := \bigcup_{l \geq 0} L^{l,k}$, where only the parameter $k$, but not the parameter
$l \leq k$ is specified, 
has been studied for example by Kolaitis and Vardi~\cite{KolaitisVardi} 
and later by Dalmau, Kolaitis, and Vardi~\cite{DalmauKolaitisVardi}.

\subsection{Datalog}
We now formally define Datalog. Our definition will be purely operational; 
for the standard semantical approach
to the evaluation of Datalog programs see~\cite{EbbinghausFlum,KolaitisVardi}.
Let $\tau$ be a relational signature.
A Datalog program (with signature $\tau$) is a finite set of \emph{rules} 
of the form $\psi \; \dlg \; \phi_1, \dots, \phi_r$, where $r \geq 0$
and where $\psi,\phi_1,\dots,\phi_r$ are atomic $\tau$-formulas. 
The formula $\psi$
is called the \emph{head} of the rule, and $\phi_1,\dots,\phi_r$ is called the \emph{body}. 
The relation symbols occurring in the head of some clause are called
\emph{intentional database predicates} (or \emph{IDBs}), 
and all other relation symbols in the clauses are 
called \emph{extensional database predicates} (or \emph{EDBs}).
A Datalog program has \emph{width $(l,k)$} if all IDBs are at most
$l$-ary, and if all rules have at most $k$ distinct variables.
A Datalog program has \emph{width~$l$} if it has width $(l,k)$ for some $k$. 

%


An \emph{evaluation} of a Datalog program $\Pi$ on a finite structure $S$ proceeds in steps $i=0,1,\dots$;
at each step $i$ we maintain a $(\tau \cup \sigma)$-structure $S^i$. The relations for the symbols 
from $\tau$ are always equal to the relations from $S$, i.e.,
for every $i \geq 0$ and every $R \in \tau$ we have 
$R^{S^{i}} = R^{S}$. 
For every relation symbol $R \in \sigma$ we have that $R^{S^{i}} \subseteq R^{S^{i+1}}$ for all $i \geq 0$. 
Initially, we start with the expansion $S^0$ of $S$ where all symbols
from $\sigma$ denote the empty relation. Now suppose that 
$R_1(u^1_1,\dots,u^1_{k_1}), \dots, R_l(u^r_1,\dots,u^r_{k_r})$ 
hold in $S^i$, and that 
$$R_0(y_1^0,\dots,y_{k_0}^0) \; \dlg \; R_1(y_1^1,\dots,y_{k_1}^1),\dots,
R_l(y^r_1,\dots,y^r_{k_r})$$
is a rule from $\Pi$, 
where $u_j^i = u_{j'}^{i'}$ if $y_{j}^i = y_{j'}^{i'}$. 
Then we add the tuple $(u^0_1,\dots,u^0_{k_0})$
to $R$ in $S^{i+1}$,
where $u^0_j = u^i_{j'}$ if and only if $y^0_j = y^i_{j'}$.
We also say that the Datalog program \emph{derives} Ê$R(u^0_1,\dots,u^0_{k_0})$ from
$R_1(u^1_1,\dots,u^1_{k_1}), \dots, R_l(u^r_1,\dots,u^r_{k_r})$.
The procedure stops if no new tuples can be derived.

On an input structure with $n$ elements a Datalog program of width $l$ can derive at most $n^l$ tuples, and it is clear that
a fixed Datalog program can be evaluated on a given structure in polynomial time in the size of the structure.

We might use Datalog programs to solve constraint satisfaction problems 
$\Csp(\Gamma)$ for a template $\Gamma$ with signature $\tau$ as follows.
Let $\Pi$ be a Datalog program whose set of EDBs is $\tau$,
and let $\sigma$ be the set of IDBs of $\Pi$.
We assume that there is one distinguished 0-ary intentional relation symbol \false. 
The program $\Pi$ is \emph{sound} for $\Csp(\Gamma)$
if every finite $\tau$-structure $S$ 
does not homomorphically map to $\Gamma$ whenever
$\Pi$ derives \false\ on $S$.
We say that $\Pi$ \emph{solves} $\Csp(\Gamma)$ if
$\Pi$ derives \false\ (i.e., adds the $0$-ary tuple to the relation for the symbol \false) on $S$ if and only if 
$S$ does not homomorphically map to $\Gamma$.

Feder and Vardi showed that deciding whether a given finite template $T$ has width $1$ is decidable (see also~\cite{DalmauPearson}).
Only recently it has been shown that the question whether
a finite structure $T$ has width $(l,k)$ is decidable as well~\cite{BoundedWidth}; 
surprisingly, it turns out that $\Csp(T)$ has width $l$, for $l \geq 2$, 
if and only if it has width $2$.

\subsection{The existential pebble game}
The existential $k$-pebble game has been introduced 
to the context of
constraint satisfaction in~\cite{KolaitisVardi,DalmauKolaitisVardi,FederVardi}. 
As in~\cite{FederVardi}, we study this game with a
second parameter, 
and first define the \emph{existential $(l,k)$-pebble game}.
The usual existential $k$-pebble game is exactly the 
existential $(k-1,k)$-pebble game in our sense. Again, the second
parameter is necessary to obtain the strongest formulations of our results.


The game is played by the players 
Spoiler and Duplicator on (possibly infinite) structures $A$ and $B$ 
of the same relational signature. 
Each player has $k$ pebbles, $p_1, \dots, p_k$ for 
Spoiler and $q_1, \dots, q_k$ for Duplicator. 
Spoiler places his pebbles on elements of $A$, 
Duplicator her pebbles on elements of $B$. 
Initially, no pebbles are placed. In each round of the game
Spoiler picks $k-l$ pebbles. 
If some of these pebbles are already placed on $A$, 
then Spoiler removes them from $A$, 
and Duplicator responds by removing the corresponding pebbles from $B$. 
Spoiler places the $k-l$ pebbles on elements of $A$, 
and Duplicator responds by placing the corresponding pebbles 
on elements of $B$.
Let $i_1, \dots, i_m$ be the indices of the pebbles that are placed on $A$ 
(and $B$) after the $i$-th round. Let $a_{i_1}, \dots, a_{i_m}$ 
$(b_{i_1}, \dots, b_{i_m})$ be the elements of $A$ ($B$) pebbled 
with the pebbles $p_{i_1}, \dots, p_{i_m}$ ($q_{i_1}, \dots, q_{i_m}$) 
after the $i$-th round. If the partial mapping $h$ from $A$ to $B$ 
defined by $h(a_{i_j})=b_{i_j}$, 
for $j \in \{1,\dots,m\}$, is not a partial homomorphism from $A$ to $B$, 
then the game is over, and Spoiler wins. 
Duplicator wins if the game continues forever.

It is convenient and customary~\cite{EbbinghausFlum} to define the
existential pebble game in terms of winning positions.

\begin{definition}\label{def:duplicator-wins}
A \emph{(positional) winning strategy for Duplicator} for
the existential $(l,k)$-pebble game on $A,B$ is a non-empty set $\cal H$ of 
partial homomorphisms from $A$ to $B$ 
such that 
\begin{itemize} 
\item $\cal H$ is closed under restrictions of its members, and 
\item for all functions $h$ in $\cal H$ with $|\dom(h)| = d \leq l$
and for all $a_1,\dots,a_{k-d} \in A$ there is an extension $h'\in \cal H$
of $h$ such that $h'$ is also defined on $a_1,\dots,a_{k-d}$.
\end{itemize}
\end{definition}

Following the usual convention we take Definition~\ref{def:duplicator-wins} as the
definition of the existential pebble game. Consequently, we shall not give a
formalization of the existential pebble game in terms of sequences of rounds
and we shall not prove the equivalence between such a formalization and Definition~\ref{def:duplicator-wins}. 
Such a formalization and equivalence proof can be 
found for a closely related game, the Ehrenfeucht--\Fresse\  game, in full formal detail in~\cite{HodgesLong} (Lemma 3.2.2); the modifications of the presentation in~\cite{HodgesLong} to the existential pebble game are straightforward. 
Instead of giving these modifications here,
we shall work directly with Definition~\ref{def:duplicator-wins}. In some of our proofs,
however, it will be convenient to think about games in terms of a
sequence of moves instead of winning positions. We have taken the liberty of describing some of those arguments in
terms of a sequence of moves of Spoiler and Duplicator with
the understanding that the argument could be easily transformed into the language of winning positions.

\subsection{Treewidth}
In the remainder of this section we define the notion of \emph{treewidth} for relational structures. As in~\cite{FederVardi}, 
we need to extend the ordinary notion of treewidth for relational structures
in such a way that we can introduce the additional parameter $l$. 

Let $0\leq l<k$ be positive integers. An \emph{$(l,k)$-tree} is defined inductively as follows:
\begin{itemize}
\item A $k$-clique is an $(l,k)$-tree
\item For every $(l,k)$-tree $G$ and for every $l$-clique induced by nodes $v_1,\dots,v_l$ in $G$, 
the graph $G'$ obtained by adding $k-l$ new nodes $v_{l+1},\dots,v_k$
to $G$ and adding edges $(v_i,v_j)$ for all $i\neq j$ with $i\in\{1,\dots,k\}$, $j\in\{l+1,\dots,k\}$
(so that $v_1,\dots,v_k$ forms a $k$-clique) is also an $(l,k)$-tree.
\end{itemize}
A \emph{partial $(l,k)$-tree} is a (not necessarily induced) subgraph of an $(l,k)$-tree. \\


\begin{definition}
Let $0\leq l<k$ and let $\tau$ be a relational signature. We say that a $\tau$-structure $S$ has \emph{treewidth at most $(l,k)$} 
if the Gaifman graph of $S$ is a partial $(l,k)$-tree.
\end{definition}

If a structure has treewidth at most $(k,k+1)$ we also say that it has
\emph{treewidth at most $k$}, and it is not difficult to see that 
these structures are precisely
the structures of treewidth at most $k$ in the sense of~\cite{KolaitisVardi}. 
It is also possible to define partial $(l,k)$-trees by using 
tree-decompositions.

\begin{definition}\label{def:treedecomp}
A \emph{tree-decomposition} of a graph $G$ is a tree $T$ such that
\begin{enumerate}
\item The nodes of $T$ are sets of nodes of $G$;
\item Every edge of $G$ is entirely contained in some node of $T$;
\item If a node $v$ belongs to two nodes $x$, $y$ of $T$ it must also be in every node in the unique path from $x$ to $y$.
\end{enumerate} 
\end{definition}

A tree-decomposition $T$ is said to be of \emph{width $(l,k)$} if every node of $T$
contains at most $k$ nodes of $G$ and the intersection of two different
nodes of $T$ has size at most $l$. 
The following is a straightforward generalization of a well-known fact for single parameter $k$,
and the proof can be obtained by adapting for instance the proof given in~\cite{Leeuwen90}. 

\begin{proposition}
A graph is a partial $(l,k)$-tree if and only if it has a tree-decomposition 
of width $(l,k)$.
\end{proposition}

It was shown in~\cite{KolaitisVardi}, Lemma 5.2, that the canonical query
for a structure $S$ of treewidth at most $k$ can be expressed in 
the logic $L^{k+1}$. 
We show an analogous statement for both parameters
$l$ and $k$.

\begin{lemma}\label{lem:query}
Let $A$ be a finite structure of treewidth at most $(l,k)$. Then the canonical query
$\phi^A$ for $A$ is logically equivalent to a sentence from $L^{l,k}$.
\end{lemma}

\proof
Let $A$ be a finite structure of treewidth at most $(l,k)$, let $G$ be its Gaifman graph, and let $T$ be a tree-decomposition 
of $G$. Let us view $T$ as a rooted tree with root $t=\{a_1,\dots,a_{k'}\}$, $k'\leq k$. We shall show by structural
induction on $T$ that there exists a formula $\phi^A(y_1,\dots,y_{k'})$ in
$L^{l,k}$ with free variables 
$y_{1},\dots,y_{k'}$ such that
for every structure $B$ and elements $b_1,\dots,b_{k'}$ in $B$ 
the following two sentences
are equivalent:
\begin{enumerate}
\item[(1)] The partial mapping from $A$ to $B$ that maps $a_i$ to $b_i$ for $1 \leq i \leq k'$ can be extended to a homomorphism from $A$ to $B$;
\item[(2)] $Q(b_1,\dots,b_{k'})$ holds in $B$.
\end{enumerate}

The base case is when the tree contains only one node $t$. 
In this case $\phi^A$ is obtained by removing the existential
quantifiers in the canonical conjunctive query of $A$. For the inductive step, let $t_1,\dots,t_m$ be the children of the root $t$ in $T$. 
Consider the $m$ subtrees $T_1,\dots,T_m$ of $T$ obtained by removing $t$.
For every $i=1,\dots,m$ we root $T_i$ at $t_i$ and consider the substructure $A_i$ of $A$ induced by the set of all nodes of $A$ 
contained in some node of $T_i$. Then, $T_i$ is a tree-decomposition of $A_i$
and the induction hypothesis provides a formula $\phi^{A_i}$ for which (1) and (2) are equivalent. Let $\phi^A(y_1,\dots,y_{k'})$ be the formula
$\bigwedge \Phi$ where $\Phi$ is the following set of formulas:
\begin{itemize}
\item[(a)] For each $i=1,\dots,m$ the set $\Phi$ contains 
the formula obtained by existentially quantifying all free variables 
$y_j$ in $\phi^{A_i}$ where $a_j \not\in t$. 
Note that the resulting formula has at most $l$ free variables.
\item[(b)] The set $\Phi$ contains the conjuncts from the canonical query of the substructure of $A$ induced by the nodes in $t$ 
(as in the base case).
\end{itemize}

To show that (1) and (2) are equivalent, let $B$ be an
arbitrary structure. By the properties
of the tree-decomposition we know that $h$ is a homomorphism from $A$ to $B$ 
that maps $a_i$ to $b_i$
if and only if for all
$i=1,\dots,m$ the restriction of $h$ to $A_i$ is a homomorphism 
from $A_i$ to $B$ 
and the restriction of $h$ to the elements of $t$ is a partial homomorphism 
from $A$ to $B$ as well. The former condition is equivalent to
the fact that the assignment $y_{a_i} \mapsto b_i$ satisfies every formula of $\Phi$ included in (a).
The latter condition is equivalent to the fact that the very same assignment satisfies the formula introduced in (b).
\qed \endproof

\section{State of the art for finite domains}
\label{sect:negative}
In this section we recall the known connection between 
 the existential pebble game, finite variable logics, and bounded treewidth duality for finite structures~\cite{KolaitisVardiDatalog,KolaitisVardi,DalmauKolaitisVardi}, and show that the connection
fails if $\Gamma$ is an arbitrary structure with an infinite domain.
The equivalence of (1) and (2) has been shown in~\cite{KolaitisVardiDatalog} (Theorem 4.8 there). The equivalence of (2) and (3) follows from results
in~\cite{DalmauKolaitisVardi}; a proof of the entire theorem will appear in~\cite{AtseriasBulatovDalmauKolaitisVardi}.

A finite relational structure $S$ is a \emph{core}
if every endomorphism of $S$ is an automorphism of $S$.
It is easy to see that every finite relational structure
is homomorphically equivalent to a core, 
and that this core is unique up to isomorphism (see e.g.~\cite{HNBook}). 

\begin{theorem}\label{thm:main}
Let $A, B$ be finite relational structures over the same signature $\tau$. 
Then the following are equivalent.
\begin{enumerate}
\item Duplicator has a winning strategy for the existential 
$k$-pebble game on $A$ and $B$;
\item All $\tau$-sentences in $\exists{\cal L}^k_{\infty\, \omega}$ 
that hold in $A$ also hold in $B$;
\item Every finite $\tau$-structure whose core has treewidth at most $k-1$ that
homomorphically maps to $A$ also homomorphically maps to $B$.
\end{enumerate}
\end{theorem}

We show that for infinite structures $B$ it is in general not true
that 3 implies 1.

\begin{proposition}
There are infinite $\tau$-structures $A$ and $B$ such that
\begin{itemize}
\item  Duplicator does not have a winning strategy for the
existential $2$-pebble game on $A$ and $B$;
\item Every finite $\tau$-structure $C$ of treewidth at most $1$
that homomorphically maps to $A$ also maps to $B$.
\end{itemize}
\end{proposition}

\begin{proof}
Let $B$ be the disjoint union
of all non-isomorphic directed paths of finite length. 
Consider $A = C_3^\rightarrow$, the directed cycle on three vertices.
Every finite $\tau$-structure $C$ of treewidth at most $1$ is a finite oriented tree,
and therefore homomorphically maps to $A$ and to $B$.
However, Spoiler clearly has a winning strategy.
After Spoiler places his first pebble, Duplicator has to place his
first pebble on a path of length $l$ in $B$.
By walking with his two
pebbles in one direction on the directed cycle $A$, Spoiler
can trap Duplicator after $l$ rounds of the game.
\qed \end{proof}

The following theorem combines results obtained in~\cite{KolaitisVardiDatalog,FederVardi,DalmauKolaitisVardi}.

\begin{theorem}\label{thm:old}
Let $\Gamma$ be a $\tau$-structure over a finite domain.
Then for every $k$ the following statements are equivalent.
\begin{enumerate}
\item There is a $(k-1,k)$-Datalog program that solves $\Csp(\Gamma)$. 
\item For all finite $\tau$-structures $A$, if Duplicator has a winning
strategy for the existential $k$-pebble game on $A$ and $\Gamma$, then $A$ 
is in $\Csp(\Gamma)$.
\item The complement of $\Csp(\Gamma)$ 
can be formulated in $\exists{\cal L}^{k}_{\infty\,\omega}$.
\item For all finite $\tau$-structures $A$, 
if every finite $\tau$-structure $C$ of treewidth at most $k$ that homomorphically
maps to $A$ also maps to $\Gamma$, then $A$ homomorphically maps to $\Gamma$.
\end{enumerate}
\end{theorem}
\begin{proof}
The equivalence between 1., 2., and 3. has been shown
in~\cite{KolaitisVardi} (Theorem 4.8 there). The equivalence between 1. and 4. is due to~\cite{FederVardi} (Theorem 23 there). 
\qed \end{proof}

Also Theorem~\ref{thm:old} fails for structures $\Gamma$ 
over an infinite domain. 
Intuitively, the reason is that
the expressive power of infinitary disjunction is relatively
larger for $\Csp(\Gamma)$ if $\Gamma$ has an infinite domain.

\begin{proposition}
There are an infinite $\tau$-structure $\Gamma$ and a finite $\tau$-structure $A$ such that 
\begin{itemize}
\item the complement of $\Csp(\Gamma)$ \emph{can} be formulated 
in $\exists{\cal L}^2_{\infty\,\omega}$ and
\item Duplicator wins the existential $2$-pebble game on $A$ and $\Gamma$, 
 but $A$ is not in $\Csp(\Gamma)$.
\end{itemize}
\end{proposition}

\begin{proof}
We choose $\Gamma$ to be $(\mathbb Q,<)$.
Duplicator wins the existential $2$-pebble game
on $C_3^\rightarrow$ and $\Gamma$, but there is no
homomorphism from $C_3^\rightarrow$ to $\Gamma$.

The complement of $\Csp(\mathbb Q,<)$ can be formulated 
in $\exists{\cal L}^2_{\infty\,\omega}$.
Let $\Phi$ be
an $\exists{\cal L}^2_{\infty\,\omega}$-sentence 
that expresses that a structure contains copies of $(\{1,\dots,n\},<)$ for
arbitrarily large $n$.
The finite directed graphs that do not homomorphically map to $(\mathbb Q,<)$ 
are precisely the directed graphs
containing a directed cycle. Clearly, $\Phi$ holds precisely on those finite directed
graphs that contain a directed cycle.
\qed \end{proof}

\section{Datalog for $\omega$-categorical structures}
\label{sect:datalog-omegacat}
The concept of $\omega$-categoricity is of central interest in model theory~\cite{HodgesLong,Oligo}. We show that many facts that are known about Datalog programs for finite structures extend to $\omega$-categorical structures.

\subsection{Countably categorical structures}
\label{sect:omegacat}
A countable structure $\Gamma$ is called \emph{$\omega$-categorical} if all 
countable models of the first-order theory 
of $\Gamma$ are isomorphic to $\Gamma$.
The following is a well-known and fundamental connection that
shows that $\omega$-categoricity of $\Gamma$ is a property of the automorphism
group of $\Gamma$, 
without reference to concepts from logic (see~\cite{HodgesLong}). 
The \emph{orbit} of an $n$-tuple $\overline a$ from $\Gamma$ 
is the set $\{\alpha(\overline a) \; | \; \alpha \text{ is an automorphism of } \Gamma\}$. 


\begin{theorem}[Engeler, Ryll-Nardzewski, Svenonius; see e.g.~\cite{HodgesLong}]\label{thm:ryll}
The following properties of a countable structure $\Gamma$ 
are equivalent:
\begin{enumerate}
\item the structure $\Gamma$ is $\omega$-categorical;
\item for each $n\geq 1$, there are finitely many orbits of $n$-tuples 
in the automorphism group of $\Gamma$;
\item for each $n \geq 1$, there are finitely many inequivalent 
first-order formulas with $n$ free variables over $\Gamma$.
\end{enumerate}
\end{theorem}

\paragraph{Examples.} An example of an $\omega$-categorical directed
graph is the set of rational numbers with the dense linear order $(\mathbb{Q}, <)$~\cite{HodgesLong}. The CSP for this structure is digraph acyclicity.

Another important example
it the \emph{universal triangle free graph $\ntriangleleft$}. 
This structure is the up to isomorphism unique countable $K_3$-free 
graph with the following \emph{extension property}: 
whenever $S$ is a subset and $T$ 
is a disjoint independent subset of the vertices in $\ntriangleleft$, then
$\ntriangleleft$ contains a vertex $v \notin S \cup T$ that is linked
to no vertex in $S$ and to all vertices in $T$.
Since the extension property can be formulated 
by an (infinite) set of first-order sentences, it follows that 
$\ntriangleleft$ is $\omega$-categorical~\cite{HodgesLong}. 
The structure $\ntriangleleft$
is called the \emph{universal} triangle free graph, 
because every other countable
triangle free graph embeds into $\ntriangleleft$.
The problem $\Csp(\ntriangleleft)$ is the problem to decide whether a given graph does not contain triangles. 
This problem is clearly polynomial-time tractable;
however, it can not be formulated
as a constraint satisfaction problem with 
a finite template~\cite{FederVardi,MadelaineStewart}.

The following lemma states an important 
property of $\omega$-categorical structures needed several times
later. The proof contains a typical proof technique for $\omega$-categorical 
structures.

\begin{lemma}\label{lem:infinst}
Let $\Gamma$ be a finite or infinite $\omega$-categorical 
structure with relational
signature $\tau$, and let $\Delta$ be a countable relational structure
with the same signature $\tau$. 
If there is no homomorphism from $\Delta$ to $\Gamma$, then
there is a finite substructure of $\Delta$
that does not homomorphically map to $\Gamma$.
\end{lemma}

\begin{proof}
Suppose every finite substructure of $\Delta$ homomorphically maps
to $\Gamma$. We show the contraposition of the lemma,
and prove the existence of a homomorphism from $\Delta$ to $\Gamma$.
Let $a_1, a_2, \dots$ be an enumeration of $\Delta$.
We construct a directed acyclic graph with finite out-degree, 
where each node lies on some level $n \geq 0$.
The nodes on level $n$
are equivalence classes of homomorphisms from the substructure of $\Delta$
induced by $a_1, \dots, a_n$ to
$\Gamma$. Two such homomorphisms $f$ and $g$ 
are equivalent, if there is an automorphism
$\alpha$ of $\Gamma$ such that $f\alpha = g$. Two equivalence classes
of homomorphisms on level $n$ and $n+1$ are adjacent, if there are 
representatives of the classes such that one is a restriction of the other.
Theorem~\ref{thm:ryll} asserts that $\Gamma$ has
only finitely many orbits of $k$-tuples, for all $k\geq 0$
(clearly, this also holds if $\Gamma$ is finite).
Hence, the constructed directed graph 
has finite out-degree. By assumption, there is
a homomorphism from 
the structure induced by $a_1, a_2, \dots, a_n$ to $\Gamma$ for all $n\geq 0$,
and hence the directed graph has vertices on all levels. 
K\"onig's Lemma asserts the existence of an infinite
path in the graph, which can be used to inductively define a homomorphism $h$ from $\Delta$ to $\Gamma$ as follows.

The restriction of $h$ to $\{a_1,\dots,a_n\}$
will be an element from the $n$-th node of the infinite path in $G$. Initially,
this is trivially true if $h$ is restricted to the empty set.
Suppose $h$ is already defined on $a_1,\dots,a_n$, for $n\geq 0$. By
construction of the infinite path, we find representatives $h_n$ and
$h_{n+1}$ of the $n$-th and the $n+1$-st element on the path such
that $h_n$ is a restriction of $h_{n+1}$. The inductive assumption
gives us an automorphism $f$ of $\Gamma$ such that $f(h_n(x))=h(x)$
for all $x \in \{a_1,\dots,a_n\}$. We set $h(a_{n+1})$ to be
$f(h_{n+1}(a_{n+1}))$. The restriction of $h$ to $a_1,\dots,a_{n+1}$
will therefore be a member of the $n+1$-st element of the infinite
path. The operation $f$ defined in this way is indeed a homomorphism from $\Delta$ to $\Gamma$. 
\qed \end{proof}

\subsection{Canonical Datalog programs}
\label{sect:canonical}
In this section we define the canonical Datalog program
of an $\omega$-categorical
structure $\Gamma$ with finite relational signature $\tau$. 
We will later prove in Section~\ref{sect:datalog}
that $\Csp(\Gamma)$ can be solved by 
an $(l,k)$-Datalog program if and only if the canonical 
$(l,k)$-Datalog program solves the problem. 

For \emph{finite} $\tau$-structures $T$ 
the canonical Datalog program for $T$ 
was defined in~\cite{FederVardi}. Our definition generalizes this definition
to $\omega$-categorical structures $\Gamma$. 
The \emph{canonical $(l,k)$-Datalog program} for $\Gamma$ 
contains an IDB for every at most $l$-ary primitive positive definable
relation in $\Gamma$. The empty $0$-ary relation serves as \false (this relation is primitive positive definable unless $\Csp(\Gamma)$ is trivial in the sense that every instance has a solution; our definition applies to non-trivial CSPs only). 
The input relation symbols are precisely the relation symbols from $\tau$.

Let $\Gamma'$ be the expansion of $\Gamma$ by all at most $l$-ary
primitive positive definable relations in $\Gamma$. It is a well-known consequence of 
Theorem~\ref{thm:ryll} that first-order expansions of  $\omega$-categorical structures, and hence in particular the structure $\Gamma'$, are also $\omega$-categorical. 
The new relations of $\Gamma'$ will be the IDBs 
and the relations that were already present in $\Gamma$ are the EDBs of the canonical Datalog program. Theorem~\ref{thm:ryll} 
also asserts that over $\Gamma'$ there is a finite number of
inequivalent formulas $\Psi(\overline x)$ of the form 
$$\big (\exists \overline y (\psi_1(\overline x, \overline y) \wedge \dots \wedge \psi_j(\overline x,\overline y))\big ) \rightarrow R (\overline x)$$ having at most $k$ variables,
where $\psi_1, \dots, \psi_j$ are atomic formulas of the 
form $R_1(\overline z_1),\dots,R_j(\overline z_j)$
for IDBs or EDBs $R_1,\dots,R_j$ and an IDB $R$.
For each of these inequivalent implications $\Psi(\overline x)$ we introduce a rule 
$$R(\overline x) \; \dlg \; R_1(\overline z_1), \dots, R_j(\overline z_j)$$
into the canonical Datalog program if 
$\forall \overline x. \Psi(\overline x)$ is valid in $\Gamma'$. In other
words, we introduce this rule if $R(\overline x)$ is implied by 
$\exists \overline y \big(\psi_1(\overline x,\overline y) \wedge \dots \wedge \psi_j(\overline x, \overline y)\big)$ in $\Gamma'$.
Since there are finitely many implications $\Psi$ that are pairwise inequivalent in $\Gamma'$,
the canonical $(l,k)$-Datalog program is finite. 

Observe that the final stage of the evaluation  
of the canonical Datalog program $\Pi$ on a given instance $S$ of $\Csp(\Gamma)$
gives rise to an instance $S'$ of $\Csp(\Gamma')$ (where $\Gamma'$ is as defined in the previous paragraph), namely the expansion of $S$
computed in the last step of the evaluation of $\Pi$ on $S$: since the IDBs of
$\Pi$ are relations from $\Gamma'$, the structure computed at the last step of the evaluation of $\Pi$ on $S$ is an instance of $\Csp(\Gamma')$. 

The following is easy to see.

\begin{proposition}\label{prop:sound}
Let $\Gamma$ be an $\omega$-categorical structure with finite relational signature. 
Then the canonical $(l,k)$-Datalog 
program for $\Gamma$ is sound for $\Csp(\Gamma)$.
\end{proposition}

\begin{proof}
We have to show that if the canonical $(l,k)$-Datalog 
program derives \false on a given instance $S$,
then $S$ is unsatisfiable. 
We claim that when the canonical
Datalog program derives $R(\bar c)$ for some tuple
$\bar c=(c_1,\dots,c_d)$ of elements of $S$, and the IDB 
$R$ has been introduced for the relation 
with the primitive positive formula $\phi(x_1,\dots,x_d)$ over $\Gamma$, then
for all homomorphisms $f$ from $S$ to $\Gamma$ we have
that $\Gamma$ satisfies $\phi(f(c_1),\dots,f(c_d))$. 
This follows by a straightforward induction over the
evaluation of canonical Datalog programs, 
using the fact that the rules of the canonical Datalog program have been introduced for valid implications in the expansion $\Gamma'$ of $\Gamma$ by all at most $l$-ary primitive positive definable relations in $\Gamma$.
Now, if the canonical $(l,k)$-program 
for $\Gamma$ derives \false on an instance $S$ of $\Csp(\Gamma)$, then this shows that there is no
homomorphism from $S$ to $\Gamma$, and hence that $S$ is unsatisfiable. 
\qed \end{proof}

\subsection{Datalog for countably categorical structures}
\label{sect:datalog}
The following theorem is the promised link between Datalog,
the existential pebble game, finite variable logics, and hypertree duality
for $\omega$-categorical structures. We present it
in its most general form with both parameters $l$ and $k$.
The assumption of $\omega$-categoricity will be used 
for the transition from item 2 to item 3 below 
(note that the canonical Datalog program is only defined for
$\omega$-categorical structures).

\begin{theorem}\label{thm:omegacatmain}
Let $\Gamma$ be a $\omega$-categorical structure with a finite relational
signature $\tau$, and let $A$ be a finite $\tau$-structure.
Then for all $l,k$ with $l \leq k$ the following statements are equivalent.
\begin{enumerate}
\item \label{item:everysound} Every sound $(l,k)$-Datalog program for CSP($\Gamma$) does not
derive $\false$ on $A$.
\item \label{item:canonical} The canonical $(l,k)$-Datalog program for $\Gamma$ does not derive \false on $A$.
\item \label{item:duplicator} Duplicator has a winning strategy for the existential $(l,k)$-pebble 
game on $A$ and $\Gamma$.
\item \label{item:inflogic} All sentences in $L^{l,k}$
that hold in $A$ also hold in $\Gamma$.
\item \label{item:treewidth} Every finite $\tau$-structure with a core of treewidth at most $(l,k)$
that homomorphically maps to $A$ also homomorphically maps to $\Gamma$.
\end{enumerate}
\end{theorem}
\begin{proof}
The implication from 1 to 2 follows from Proposition~\ref{prop:sound}.

To show that~\ref{item:canonical} implies~\ref{item:duplicator},
we define a winning strategy for Duplicator as follows.
Let $\Gamma'$ be the expansion of $\Gamma$
by all at most $l$-ary primitive positive definable relations,
and let $A'$ be the instance of $\Csp(\Gamma')$ computed
by the canonical $(l,k)$-Datalog program for $\Gamma$
on input $A$.
Then the strategy for Duplicator contains all those partial mappings $f \colon A \rightarrow \Gamma$ 
with domain $D$ of size at most $k$ such that
for every relation $R(x_1,\dots,x_d)$ that holds in $A'$ on elements $x_1,\dots,x_d \in D$,
the tuple $(f(x_1),\dots,f(x_d))$ belongs to $R$ in $\Gamma'$.

By construction, $\mathcal H$ contains only partial homomorphisms and
is non-empty (since \false
is not derived, $\mathcal H$ contains the partial mapping with the
empty domain). We shall prove that
$\mathcal H$ has the $(l,k)$-extension property, and omit the easier proof that $\mathcal H$ is closed
under restrictions. Let $h$
be a function with domain
$v_1,\dots,v_{l'}$ of size at most $l$ and let
$D=\{v_1,\dots,v_{l'},v_{l'+1},\dots,v_{k'}\}$ be a superset of
$\{v_1,\dots,v_{l'}\}$ of size at most $k$. 
Let $T$ be the $k'$-ary relation that
contains all those tuples $(b_1,\dots,b_{k'})\in
D_{\Gamma}^{k'}$ such that $(b_{i_1},\dots,b_{i_r}) \in R^{\Gamma'}$
for every $R(v_{i_1},\dots,v_{i_r})$
 derived in $A'$ on variables $v_{i_1},\dots,v_{i_r}$ from $D$.
Consider the following rule with variables $x_1,\dots,x_{k'}$ 
of the canonical Datalog program.
The body of the rule contains for each IDB $R$ and
for all tuples $(v_{i_1},\dots,v_{i_r}) \in R^{A'}$
such that $v_{i_1},\dots,v_{i_r} \in D$
 the atomic predicate $R(x_{i_1},\dots,x_{i_r})$.
The head of the rule is $S(x_1,\dots,x_{l'})$ where
$S^{\Gamma'}$ is the projection of the relation $T$ to the first $l'$ arguments.
The instantiation $x_i \rightarrow v_i$, $i=1,\dots,k'$, allows
to derive $S(v_1,\dots,v_{l'})$ by this rule, and, by the definition of $\mathcal H$,
$(h(v_1),\dots,h(v_{l'}))$ belongs to $S^{\Gamma'}$.
By the definition of $S^{\Gamma'}$, there exist
$b_{l'+1},\dots,b_{k'}$ such that
$(h(v_1),\dots,h(v_{l'}),b_{l'+1},\dots,b_{k'})$ belongs to $T$. Hence,
if we extend $h$ by $v_i \rightarrow b_i$ for $i$ from $l'+1,\dots,k'$ we obtain the desired function.

Next, we show the implication from~\ref{item:duplicator} 
to~\ref{item:inflogic}.
The proof closely follows the corresponding proof for finite structures
given in~\cite{KolaitisVardi}, with the important difference that
we have both parameters $l$ and $k$ in our proof, whereas
previously the results have only been stated with the parameter $k$.

Suppose Duplicator has a winning strategy $\cal H$
for the existential $(l,k)$-pebble game on $A$ and $\Gamma$.
Let $\phi$ be a $\tau$-sentence from $L^{l,k}$ that holds in $A$.
We have to show that $\phi$ also holds in $\Gamma$. 
For that, we prove by 
induction on the syntactic structure of $L^{l,k}$ 
formulas that 

\emph{
if $\psi(v_1,\dots,v_m)$ is an $L^{l,k}$ formula
that is an $l$-bounded conjunction or has at most $l$ free variables (i.e., $m \leq l$),
then for all $h \in \cal H$ and all elements $a_1,\dots,a_m$ from the domain of $h$, if $A$ satisfies $\psi(a_1,\dots,a_m)$, then
$\Gamma$ satisfies $\psi(h(a_1),\dots,h(a_m))$.}

Clearly, choosing $m=0$, this implies that $\phi$ holds in $\Gamma$.
The base case of the induction is obvious, since atomic
formulas are preserved under homomorphisms. Next, 
suppose that $\psi(v_1,\dots,v_m)$ is an $l$-bounded conjunction
of a set of formulas $\Psi$.
Then each formula in $\Psi$ either has at most $l$ free variables,
or is quantifier-free. In both cases we can use the inductive hypothesis,
and the inductive step follows directly.

Assume that the formula $\psi(v_1,\dots,v_m)$ is of the form
$\exists u_1,\dots,u_n. \, \chi(v_1,\dots,v_m,u_1,\dots,u_n)$.
Since $\psi$ is from $L^{l,k}$, 
we know that $n+m \leq k$.
If $m > l$, there is nothing to show. Otherwise,
we choose $\chi$ and $n$ such that $n$ is largest possible. 
Therefore, $\chi$ is either
an $l$-bounded conjunction or
an atomic formula. 
We will use the inductive hypothesis for the formula $\chi(v_1,\dots,v_m,u_1,\dots,u_n)$.
Let $h$ be a homomorphism in $\cal H$. We have to show that if $a_1,\dots,a_m$
are arbitrary elements from the domain of $h$ such that
$A$ satisfies $\psi(a_1,\dots,a_m)$, 
then $\Gamma$ satisfies $\psi(h(a_1),\dots,h(a_m))$. 


Since $A$ satisfies $\exists u_1,\dots, u_n. \chi (a_1,\dots,a_m)$,
there exist $a_{m+1},\dots,a_{m+n}$
such that $A$ satisfies $\chi(a_1,\dots,a_m,a_{m+1},\dots,a_{m+n})$.
Consider the restriction $h^*$ of $h$ to the subset $\{a_1,\dots,a_m\}$
of the domain of $h$. Because of the first property of winning strategies
$\cal H$, the homomorphism $h^*$ is in $\cal H$. 
Since $m \leq l$, we can apply the forth property of $\cal H$
to $h^*$ and $a_{m+1},\dots,a_{m+n}$, and there are $b_1,\dots,b_n$
such that the extension $h'$ of $h^*$ with domain 
$\{a_1,\dots,a_{m+n}\}$ that maps $a_{m+i}$ to $b_i$ is in $\cal H$. 
By applying the induction hypothesis 
to $\chi(v_1,\dots,v_m,u_1,\dots,u_n)$ and to $h'$, we infer that 
$\Gamma$ satisfies $\chi(h'(a_1),\dots,h'(a_{m+n}))$, 
and hence $\Gamma$ satisfies $\psi(h(a_1),\dots,h(a_m))$.


\ref{item:inflogic} implies~\ref{item:treewidth}. Let $T$ be a finite 
$\tau$-structure $T$ whose core $T'$ has treewidth at most $(l,k)$ 
such that $T$ homomorphically maps to $A$. 
By Lemma~\ref{lem:query} there exists an $L^{l,k}$-sentence $\phi$ 
such that $\phi$ holds in a structure $B$ if and only if $T'$
homomorphically maps to $B$. In particular, $\phi$ must hold in $A$.
Then $4$ implies that $\phi$ holds in $\Gamma$, and therefore
$T'$ homomorphically maps to $\Gamma$. But then we can compose the homomorphism from $T$ to $T'$ and the homomorphism from $T'$ to $\Gamma$ to obtain the desired homomorphism from $T$ to $\Gamma$.


We finally show that~\ref{item:treewidth} implies~\ref{item:everysound}. 
Assume~\ref{item:treewidth}, and suppose for contradiction
that there is a sound $(l,k)$-Datalog program $\Pi$ for $\Gamma$
that derives $\false$ on $A$.
The idea is to use the `derivation tree of \false' to 
construct a $\tau$-structure $S$ of treewidth at most $(l,k)$ 
that homomorphically maps
to $A$, but not to $\Gamma$.
The construction proceeds by induction over the evaluation of
$\Pi$ on $A$. 
Suppose that $R_0(y_1^0,\dots,$
$y_{k_0}^0)$ is an atomic formula derived by $\Pi$ on $A$ from previously derived atomic formulas
$R_1(y_1^1,\dots,y_{k_1}^1),\dots,R_s(y^s_1,\dots,y^s_{k_s})$.
We will prove that there exists a structure $S_0$
with distinguished vertices $v_1^0,\dots,v_{k_0}^0$ and an $(l,k)$-tree
$G_0$ such that
\begin{enumerate}
\item the Gaifman graph of $S_0$ is a (not necessarily
induced) subgraph of $G_0$, 
\item $v_1^0,\dots,v_{k_0}^0$ induce a
clique in $G_0$, 
\item there is a homomorphism from $S_0$ to $A$ 
that maps $v_i^0$ to $y_i^0$ for every $1\leq i\leq k_0$, and 
\item
the program $\Pi$ derives $R_0(v_1^0,\dots,v_{k_0}^0)$ on $S_0$.
\end{enumerate}

Let $i \in \{1,\dots,s\}$. If $R_i$ is an IDB, then let $S_i$,
$v_1^i,\dots,v_{k_i}^i$, and $G_i$ be given by the inductive hypothesis.
If $R_i$ is an EDB, we create fresh vertices $v^i_1,\dots,v^i_{k_i}$,
and define $S_{i}$ to be the following structure with vertices $v^i_1,\dots,v^i_{k_i}$. 
The relation
$R_i$ in $S_{i}$ equals $\{(v^i_1,\dots,v^i_{k_i})\}$, and all other relations in $S_{i}$ are empty. Clearly, $\{v^i_1,\dots,v^i_{k_i}\}$ induces
a clique in the Gaifman graph of $S_i$, and the Gaifman graph of $S_i$ 
is a partial $(l,k)$-tree.

Now, the structure $S_{0}$ has
the distinguished vertices
$v^0_1,\dots,v^0_{k_0}$, and is obtained from the $\tau$-structures $S_{1},\dots,S_{s}$ as follows.
We start from the disjoint union of $S_{1},\dots,S_{s}$.
When $y^i_j = y^r_s$ for $i,r \in \{0,\dots, s\}$, $j \in \{1,\dots,k_i\}$, and $s \in \{1,\dots,k_r\}$, then we
 identify $v^i_j$ and $v^r_s$.
To define
$G_0$ we form a disjoint union of $G_1,\dots,G_s$ and the isolated
nodes $v_1^0,\dots,v_{k_0}^0$, and do the same node identifications as
before. We finally add an edge for every pair of distinct vertices in $v_1^0,\dots,v_{k_0}^0$. The resulting graph, $G_0$,
satisfies the requirements of the claim.
Observe that since $\Pi$ derives 
$R_1(v_1^1,\dots,v_{k_1}^1),
\dots, R_s(v^s_1,\dots,v^s_{k_s})$ on $S_{0}$
by inductive assumption, it also derives
$R_0(v_1^0,\dots,v_{k_0}^0)$ on $S_{0}$.

In this fashion we proceed for all inference steps of the Datalog program. 
Let $S$ be the resulting
structure for the final derivation of false. It has treewidth at most
$(l,k)$, and maps to $S$, but does not map to 
$\Gamma$, since $\Pi$
(which is sound) derives also \false\ on $S$.
\qed \end{proof}

\subsection{Application to constraint satisfaction}
We discuss an important consequence of Theorem~\ref{thm:omegacatmain} 
with many concrete applications: 
we prove that $\Csp(\Gamma)$ for $\omega$-categorical
$\Gamma$ is tractable if the input is restricted to instances
of treewidth at most $(l,k)$. In fact, for the tractability result 
we only have to require
that the \emph{cores} of the input structures have bounded treewidth.
The statement where we only require that the core of
input structures
has treewidth at most $(l,k)$ is considerably stronger (also see~\cite{GroheJournal});
the corresponding statement for finite structures and single parameter
$k$ has been observed in~\cite{DalmauKolaitisVardi}.

\begin{corollary}\label{cor:twdatalogcan}
Let $\Gamma$ be an $\omega$-categorical structure with finite relational
signature $\tau$. Then every instance $A$ of $\Csp(\Gamma)$
whose core has treewidth at most $(l,k)$ can be solved in polynomial time by 
the canonical $(l,k)$-Datalog program.
\end{corollary}
\begin{proof}
It is clear that an $(l,k)$-Datalog program can be evaluated on
a (finite) instance $A$ of $\Csp(\Gamma)$ in polynomial time.
If the canonical $(l,k)$-Datalog program derives $\false$
on $A$, then, because the canonical Datalog program is always sound,
the instance $A$ is not homomorphic to $\Gamma$.
Now, suppose that the canonical Datalog program does not derive
\false on a finite structure $A$ whose core has treewidth at most $(l,k)$. Then, 
by Theorem~\ref{thm:omegacatmain}, every $\tau$-structure 
whose core has treewidth at most $(l,k)$ that homomorphically 
maps to $A$ also homomorphically maps to $\Gamma$. This
holds in particular for $A$ itself, and hence $A$
is homomorphic to $\Gamma$.
\qed \end{proof}

The following direct consequence of Theorem~\ref{thm:omegacatmain} yields other characterizations of bounded Datalog width.

\begin{theorem}\label{thm:main2}
Let $\Gamma$ be a $\omega$-categorical structure with a finite relational
signature $\tau$. Then for all $l,k$ with $l \leq k$ the following statements are equivalent.
\begin{enumerate}
\item There is an $(l,k)$-Datalog program that solves $\Csp(\Gamma)$.
\item The canonical $(l,k)$-Datalog program solves $\Csp(\Gamma)$.
\item For all finite $\tau$-structures $A$, 
if Duplicator has a winning strategy for the existential $(l,k)$-pebble 
game on $A$ and $\Gamma$, then $A$ is in $\Csp(\Gamma)$. 
\item For all finite $\tau$-structures $A$, if all sentences in $L^{l,k}$
that hold in $A$ also hold in $\Gamma$, 
then $A$ homomorphically maps to $\Gamma$.
\item For all finite $\tau$-structures $A$, if every finite $\tau$-structure $S$ of 
treewidth at most $(l,k)$ that
homomorphically maps to $A$ also homomorphically maps to $\Gamma$, then
$A$ homomorphically maps to $\Gamma$.
\item There is a set $\cal N$ of finite structures of treewidth at most 
$(l,k)$ such that every finite $\tau$-structure $A$ is homomorphic to $\Gamma$ 
if and only if no structure in $\cal N$ is homomorphic to $A$.
\end{enumerate}
\end{theorem}

\begin{proof}
To prove the implication from 1 to 2,
suppose that an $(l,k)$-Datalog program $\Pi$ solves $\Csp(\Gamma)$,
and let $S$ be an instance of $\Csp(\Gamma)$. If the canonical $(l,k)$-Datalog
program derives \false on $S$, then by Proposition~\ref{prop:sound} the
structure $S$ is not homomorphic to $\Gamma$. 
Otherwise, since $\Pi$ is sound, the implication from 2 to 1 in 
Theorem~\ref{thm:omegacatmain} shows that the canonical $(l,k)$-Datalog
program does not derive \false on $S$ as well. Hence, the canonical Datalog
program solves $\Csp(\Gamma)$.

The implications $2 \Rightarrow 3 \Rightarrow 4 \Rightarrow 5 \Rightarrow 1$
are straightforward 
consequences of Theorem~\ref{thm:omegacatmain}.

To show that 5 implies 6, let $\cal N$ be the set
of all those structures of treewidth at most $(l,k)$ that does not homomorphically
map to $\Gamma$. Let $A$ be a finite $\tau$-structure.
If $A$ homomorphically maps to $\Gamma$, then clearly 
there is no structure $C$ 
in $\cal N$ that maps to $A$, because then $C$ would also map to $\Gamma$,
a contradiction to the definition of $\cal N$. 
Conversely, suppose that no structure in $\cal N$ homomorphically maps
to $A$. In other words, every structure that homomorphically maps to $A$
also maps to $\Gamma$. Using 5, this implies that $A$ homomorphically
maps to $\Gamma$.

Finally, 6 implies 5. Let $\cal N$ be such that it satisfies the conditions of item 6. 
It follows that all structures  in $\cal N$ do not map homomorphically to $\Gamma$.
Let $A$ be a finite $\tau$-structure such that every finite $\tau$-structure $S$
of treewidth at most $(l,k)$ that homomorphically maps to $A$ also
homomorphically maps to $\Gamma$. In particular, no structure in $\cal N$
homomorphically maps to $A$. Therefore, $A$ homomorphically maps to $\Gamma$.
\qed \end{proof}

\section{1-Datalog, MMSNP, and constraint satisfaction}
\label{sect:datalogascsp} 
A Datalog program of width one accepts a class of
structures that can be described by a sentence of a fragment of
existential second order logic called \emph{monotone monadic SNP
without inequalities (MMSNP)}. We show that every problem in MMSNP that is closed under disjoint unions 
can be formulated as the constraint satisfaction problem for an $\omega$-categorical template.
It follows that for every infinite structure $\Gamma$ with
finite relational signature, if $\Csp(\Gamma)$ has
Datalog width one, then $\Csp(\Gamma)$ can also be formulated as a constraint satisfaction problem with an $\omega$-categorical template. 

An \emph{SNP sentence} is an existential second-order sentence with
a universal first-order part. The first order part might contain the
existentially quantified relation symbols and additional relation
symbols from a given signature $\tau$ (the \emph{input} relations).
We shall assume that SNP formulas are written in \emph{negation
normal form}, i.e., the first-order part is in prenex normal form, and the quantifier-free part is in conjunctive
normal form where each disjunction is written as a negated
conjunction of positive and negative literals. It is well-known that every first-order formula is logically equivalent to a formula of this form. 
SNP sentences can be used to \emph{describe} computational problems in the sense that an SNP sentence $\Phi$ is valid on a structure $A$ if and only if $A$ is a yes-instance of the respective computational problem. 
The \emph{class} SNP
consists of all problems on relational $\tau$-structures that can be described by an SNP sentence.

The class \emph{MMSNP}, defined by Feder and Vardi, is the class of
problems that can be described by an SNP sentence $\Phi$ that satisfies three additional requirements:
\begin{itemize}
\item the existentially quantified relations in $\Phi$
are \emph{monadic}, that is, unary, 
\item $\Phi$ is \emph{monotone}, i.e., every input relation symbol occurs \emph{negatively}
in $\Phi$, and 
\item $\Phi$ does not contain inequalities. 
\end{itemize}
Every problem in MMSNP is equivalent under randomized
Turing reductions to a constraint satisfaction problem
with a finite template~\cite{FederVardi}; a deterministic reduction
was announced by Kun~\cite{Kun}. 
It is easy to see
that MMSNP contains
all constraint satisfaction problems with finite templates. Thus,
MMSNP has a complexity dichotomy (meaning that every problem in MMSNP is polynomial-time solvable or NP-complete) 
if and only if the class of all finite-domain CSPs has a dichotomy.

It has already been observed by Feder and Vardi~\cite{FederVardi} that $(1,k)$-Datalog is contained in the class MMSNP.  
For a proof, introduce an existentially quantified unary predicate for each of
the unary IDBs in the Datalog program. It is then straightforward to
translate the rules of the Datalog program into universal first-order formulas
with at most $k$ first-order variables. 

We now want to prove that
every problem in MMSNP can be formulated as a constraint
satisfaction problem with a countably categorical template. In full
generality, this cannot be true because
constraint satisfaction problems are always closed under disjoint union. A simple example of
an MMSNP problem not closed under disjoint union is the one defined
by the formula $\forall x,y \; \neg(P(x)\wedge Q(x))$. Hence, we
shall assume that we are dealing with a problem in
MMSNP that is {\em closed under disjoint union}.

To prove the claim under this assumption, we need a recent model-theoretic result of Cherlin, Shelah
and Shi~\cite{CherlinShelahShi}. Let $\cal N$ be a finite set of
finite structures with a relational signature $\tau$. In this paper,
a $\tau$-structure $\Delta$ is called \emph{$\cal N$-free} if there
is no homomorphism from any structure in $\cal N$ to $\Delta$. A
structure $\Gamma$ in a class of countable structures $\cal C$ is
called \emph{universal} for $\cal C$, if it contains all structures
in $\cal C$ as an induced substructure.

\begin{theorem}[of \cite{CherlinShelahShi}]\label{thm:universal}
Let $\cal N$ be a finite set of finite connected $\tau$-structures.
Then there is an $\omega$-categorical structure $\Delta$
that is universal for the class of all countable $\cal N$-free
structures.
\end{theorem}

Cherlin, Shelah and Shi proved this statement for (undirected)
graphs, but the proof does not rely on this assumption on the
signature, and works for arbitrary relational signatures. The
statement in its general form also follows from a result
in~\cite{Covington}. We use the $\omega$-categorical structure $\Delta$ to
prove the following.

\begin{theorem}\label{thm:mmsnp}
Every problem in MMSNP that is closed under disjoint unions can
be formulated as $\Csp(\Gamma)$ with an $\omega$-categorical template
$\Gamma$.
\end{theorem}

\begin{proof}
Let $\Phi$ be a MMSNP sentence with signature $\tau$,
written in negation normal form, whose class $\cal M$ of finite models is closed under disjoint unions. We have
to find an $\omega$-categorical $\tau$-structure $\Gamma$ such that
$\cal M$ equals $\Csp(\Gamma)$. 
Let $P_1, \dots, P_k$ be the
existential monadic predicates in $\Phi$. 
For each existential monadic relation
$P_i$ we introduce a relation symbol $P'_i$, and replace negative
literals of the form $\neg P_i(x)$ in $\Phi$ by $P_i'(x)$.
We shall denote the formula obtained after this transformation by
$\Phi'$. Let $\tau'$ be the signature containing the input relations
from $\tau$, the existential monadic relations $P_i$, and the
symbols $P_i'$ for the negative occurrences of the existential
relations. We define ${\cal N}$ to be the set of $\tau'$-structures
containing for each clause $\neg(L_1\wedge\dots\wedge L_m)$ in
$\Phi'$ the canonical database~\cite{ChandraMerlin} of
$(L_1\wedge\dots\wedge L_m)$. We shall use the fact that a
$\tau'$-structure $S$ satisfies a clause $\neg(L_1\wedge\dots\wedge
L_m)$ if and only if the the canonical database of
$(L_1\wedge\dots\wedge L_m)$ is not homomorphic to $S$.



We can assume without loss of generality that $\Phi$ is minimal in
the sense that if we remove a literal from some of the clauses the
formula obtained is inequivalent. We shall show that then all
structures in $\cal N$ are connected. Let us suppose that this is
not the case. Then there is a clause $C$ in $\Phi$ that corresponds to a
non connected structure in ${\cal N}$. The clause $C$ can be written
as $\neg(E\wedge F)$ where the set $X$ of variables in $E$ and the
set $Y$ of variables in $F$ do not intersect. Consider the formulas
$\Phi_E$ and $\Phi_F$ obtained from $\Phi$ by replacing $C$ by $\neg
E$ and $C$ by $\neg F$, respectively. By minimality of $\Phi$ there
is a structure $M_E$ that satisfies $\Phi$ but not $\Phi_E$, and
similarly there exists a structure $M_F$ that satisfies $\Phi$ but
not $\Phi_F$. By assumption, 
the disjoint union $M$ of $M_E$ and $M_F$ satisfies
$\Phi$. Then there exists a $\tau''$-expansion $M''$ of $M$ where
$\tau'' := \tau\cup\{P_1,\dots,P_k\}$ that satisfies the first-order part of $\Phi$. Consider the substructures $M''_E$ and $M''_F$ of
$M''$ induced by the vertices of $M_E$ and $M_F$. We have that
$M''_E$ does not satisfy the first-order part of 
$\Phi_E$ (otherwise $M_E$ would satisfy
$\Phi_E$). Consequently, there is an assignment $s_E$ of the
universal variables that falsifies some clause. This clause must
necessarily be $\neg E$ (since otherwise $M''$ would not satisfy the
first-order part of $\Phi$). By similar reasoning we can infer that
there is an assignment $s_F$ of the universal variables of $\Phi$ to
elements of $M_F$ that falsifies $\neg F$. Finally, fix any
assignment $s$ that coincides with $s_E$ over $X$ and with $s_F$
over $Y$ (such an assignment exists because $X$ and $Y$ are
disjoint). Clearly, $s$ falsifies $C$ and $M$ does not satisfy
$\Phi$, a contradiction. Hence, we shall assume that every structure
in $\cal N$ is connected.

Then Theorem~\ref{thm:universal} asserts the existence of an $\cal N$-free $\omega$-categorical $\tau'$-structure $\Delta$ that is
universal for all $\cal N$-free structures. We use $\Delta$ to
define the template $\Gamma$ for the constraint satisfaction
problem. To do this, let $\Delta'$ be the restriction of $\Delta$ to those elements
that have the property that for all existential monadic predicates $P_i$ either $P_i$ or $P_i'$ holds (but not both $P_i$ and $P_i'$).
Let $\Gamma$ be the reduct of $\Delta'$
that only contains the input relations from $\tau$.
It is well-known (see e.g. Theorem 7.3.8 in~\cite{HodgesLong}) hat reducts and first-order
restrictions of $\omega$-categorical structures are again
$\omega$-categorical.
Hence, $\Gamma$ is $\omega$-categorical.

We claim that a $\tau$-structure $S$ satisfies $\Phi$ if and only
if $S$ homomorphically maps to $\Gamma$. 
Suppose there is a homomorphism
$h$ from $S$ to $\Gamma$. Let $S'$ be the $\tau'$-expansion of
$S$ such that for each $i \in \{1,\dots,k\}$ the relation 
$P_i(x)$ holds in $S'$ if and
only if $P_i(h(x))$ holds in $\Delta'$, and $P'_i(x)$ holds in $S'$
if and only if $P'_i(h(x))$ holds in $\Delta'$. Clearly, $h$ defines
a homomorphism from $S'$ to $\Delta'$ and also from $S$ to $\Delta$. In
consequence, none of the structures from $\cal N$ maps to $S'$. Hence, the $\tau''$-reduction of $S'$ satisfies all the
clauses of the first-order part of $\Phi$ and hence $S$ satisfies
$\Phi$.


Conversely, let $S$ be a structure satisfying $\Phi$. Thus,
there exists a $\tau'$-expansion $S'$ of $S$ that satisfies the
first-order part of $\Phi'$ and where for every element $x$
exactly one of $P_i(x)$ or $P'_i(x)$ holds.
Clearly,
no structure in $\cal N$ is homomorphic to the expanded structure,
and by universality of $\Gamma$ the $\tau'$-structure
$S'$ is an induced substructure of $\Delta$. Since for every point
of $S'$ exactly one of $P_i$ and $P_i'$ holds, $S'
$ is also an
induced substructure of $\Delta'$. Therefore, $S$ is homomorphic
to $\Gamma$. This completes the proof. \qed
\end{proof}

In particular, we proved the following.

\begin{corollary}\label{cor:datalogadcsp}
Every problem in $(1,k)$-Datalog that is closed under disjoint
unions can be formulated as a constraint satisfaction problem with
an $\omega$-categorical template.
\end{corollary}

\begin{example}
The following computational problem is an example of a CSP in MMSNP that cannot be described with a finite template~\cite{MadelaineStewartSicomp} and that is not
in $(l,k)$-Datalog for all $1 \leq l \leq k$. We are given a finite graph $S$, and we want to
test whether we can partition the vertices of $S$ into two parts such that each part is triangle-free. 
It is easy to formulate this problem in MMSNP. Hence,
Corollary~\ref{cor:datalogadcsp} implies that it can also be formulated as a CSP with an
$\omega$-categorical template. To illustrate, we
describe such a template explicitly: 
Take two copies $C_1$ and $C_2$ of $\ntriangleleft$, and add an
undirected edge between all vertices in $C_1$ and all vertices in $C_2$.
The corresponding CSP is NP-hard~\cite{Achlioptas}.
\qed \end{example}

\section{Bounded width}
\label{sect:bounded}
In this section we characterize some families of 
$\omega$-categorical templates
whose CSPs have bounded width.
Our results generalize known algebraic characterizations
of Datalog width for constraint satisfaction with
finite templates. 
However, not all results remain valid for infinite
templates: it is well-known~\cite{FederVardi} that the
constraint satisfaction of a finite
template has Datalog width one if and only if the so-called 
\emph{arc-consistency procedure} solves the problem.
This is no longer true for infinite templates. We characterize
both width one and the expressive power of the arc-consistency procedure
for infinite $\omega$-categorical templates, and present an
example that shows that the two concepts are different.
We also present an algebraic characterization of \emph{strict width $l$}, a concept introduced by Feder and Vardi~\cite{FederVardi}.

\subsection{Width zero}
An example of a template whose constraint satisfaction problem has width 0 is
the universal triangle-free graph $\ntriangleleft$. 
Since there is a primitive positive
sentence that states the existence of a triangle in a graph,
and since every graph without a triangle is homomorphic to $\ntriangleleft$,
there is a Datalog program of width 0 
that solves $\Csp(\ntriangleleft)$.
In general, it is easy to see that 
a constraint satisfaction problem has width 0 if and only if
there is a finite set of \emph{homomorphic obstructions} for $\Csp(\Gamma)$, i.e.,
a finite set $\cal N$ of finite $\tau$-structures such that 
every finite $\tau$-structure $A$ is homomorphic to $\Gamma$ if and
only if no substructure in $\cal N$ is homomorphic to $A$.

When $\Gamma$ is a structure with finite relational signature $\tau$, we say that $\Csp(\Gamma)$ is \emph{first-order definable}
if there exists a first-order $\tau$-sentence $\Phi$ such that a finite $\tau$-structure $S$ homomorphically maps to $\Gamma$ if and only if $S$ satisfies $\Phi$. 
It turns out that $\Csp(\Gamma)$ is first-order definable
if and only if it has width 0. This can be seen as a reformulation of Rossman's theorem~\cite{Rossman08},
which says that a first-order sentence $\phi$ is equivalent to an existential positive sentence if and only if the class of finite models of $\phi$ is closed under homomorphisms.
For finite templates a characterization of first-order definable 
constraint satisfaction problems has been obtained 
in~\cite{LLT} building on work in~\cite{Atserias,NesetrilTardif}.
Our discussion is summarized by the following theorem.

\begin{theorem}
For every (not necessarily $\omega$-categorical)
template $\Gamma$ the following are equivalent.
\begin{enumerate}
\item $\Csp(\Gamma)$ has a finite obstruction set;
\item $\Csp(\Gamma)$ has Datalog width 0;
\item $\Csp(\Gamma)$ is first-order definable.
\end{enumerate}
Moreover, if $\Csp(\Gamma)$ is first-order definable
we can always find an $\omega$-categorical
structure $\Gamma'$ that has the same constraint satisfaction problem
as $\Gamma$.
\end{theorem}
\begin{proof}
The equivalence between $1.$ and $2.$ has been discussed above. For the equivalence of $2$ and $3$, note that
the complement of a CSP is closed under homomorphisms.
Hence, Rossman's theorem implies that
a CSP with an arbitrary infinite template has Datalog width 0 if and only if it is first-order definable.
The last part of the statement is a special case of Corollary~\ref{cor:datalogadcsp}. \qed
\end{proof}

\subsection{Width one}
Let $\Gamma$ be an $\omega$-categorical structure with relational
signature $\tau$,
and $\Pi$ be the canonical $(1,k)$-Datalog program for $\Gamma$,
for some $k \geq 1$. 
By Corollary~\ref{cor:datalogadcsp}, the class of $\tau$-structures
accepted by $\Pi$ is itself a CSP
with an $\omega$-categorical template, which we denote 
by $\Gamma(1,k)$. 

\begin{theorem}
Let $\Gamma$ be $\omega$-categorical. 
Then $\Csp(\Gamma)$ 
can be solved by a $(1,k)$-Datalog program
if and only if 
there is a homomorphism from $\Gamma(1,k)$ 
to $\Gamma$.
\end{theorem}

\begin{proof}
Let $\Pi$ be the canonical $(1,k)$-Datalog program
of $\Gamma$. Suppose first that there is a homomorphism from
$\Gamma(1,k)$ to $\Gamma$. We show that $\Pi$ solves $\Csp(\Gamma)$. 
Let $A$ be an instance
of $\Csp(\Gamma)$. If $\Pi$ accepts $A$, then $A$ homomorphically maps to
$\Gamma(1,k)$, and therefore also to $\Gamma$. 
Otherwise, if $\Pi$ does not accept $A$, then $A$ does not
map to $\Gamma$ since $\Pi$ is sound (Proposition~\ref{prop:sound}). 

For the opposite implication, suppose that there is
a width $(1,k)$-Datalog program that solves $\Csp(\Gamma)$. By Theorem~\ref{thm:main2}, the program $\Pi$ also
solves $\Csp(\Gamma)$. To show that $\Gamma(1,k)$ 
homomorphically maps to $\Gamma$, it suffices by 
Lemma~\ref{lem:infinst} to show that every finite substructure $A$ of the countable structure $\Gamma(1,k)$ homomorphically maps to $\Gamma$. 
Every finite substructure $A$ of $\Gamma(1,k)$ is in particular homomorphic to $\Gamma(1,k)$, 
and thus accepted by $\Pi$. Since $\Pi$ solves $\Csp(\Gamma)$, $A$ homomorphically maps to $\Gamma$. 
\qed \end{proof}


\subsection{Arc-consistency}
The \emph{arc-consistency procedure (AC)} is an algorithm for constraint
satisfaction problems that is intensively studied in Artificial Intelligence
(which is sometimes also called \emph{hyper-arc consistency} or \emph{generalized arc consistency} to stress the fact that it can also deal
with constraints of arity larger than two).
It can be described
as the subset of the canonical Datalog program of
width one that consists of all rules with bodies containing 
at most one non-IDB.
For finite templates $T$ it is known
that the arc-consistency procedure solves $\Csp(T)$ if and only
if $\Csp(T)$ has width one~\cite{FederVardi}. For infinite structures,
this is no longer true: consider for instance $\Csp(\ntriangleleft)$,
which has width 0, but cannot be solved by the arc-consistency procedure.
The reason is that the width one canonical Datalog program for $\ntriangleleft$ has no non-trivial unary predicates, and we thus have to consider at least three relations in the input 
to infer that the input contains a triangle.

The following concept is crucial to understand the power of the arc-consistency procedure. Let $\Gamma$ be an $\omega$-categorical
structure with finite relational signature $\tau$.
Since $\Gamma$ is $\omega$-categorical, there is only
a finite number of primitive positive definable nonempty sets
$O_1, \dots, O_n$. 
We define the \emph{definable subset structure}
of $\Gamma$, which is the finite
relational $\tau$-structure with domain $\{O_1, \dots, O_n\}$ where a $k$-ary relation $R$ from $\tau$ holds on $O_{i_1}, \dots, O_{i_k}$ 
iff for every $j \in \{1,\dots,k\}$ and every vertex $v_j$ in the orbit $O_{i_j}$ there 
are vertices $v_1, \dots, v_{j-1}, v_{j+1}, \dots, v_k$ from $O_{i_1}, \dots, 
 O_{i_{j-1}}, O_{i_{j+1}}, \dots, O_{i_k}$, respectively,
such that $R$ holds on $v_1, \dots, v_k$ in $\Gamma$.

\begin{lemma}
Let $\Gamma$ be an $\omega$-categorical structure with finite relational signature $\tau$. Then
for every instance $S$ of $\Csp(\Gamma)$ the following
two statements are equivalent:
\begin{enumerate}
\item The arc-consistency procedure $\Pi$ for $\Gamma$ does not derive \emph{false} on
instance $S$.
\item $S$ is homomorphic to the definable subset structure of $\Gamma$.
\end{enumerate}
\end{lemma}
\begin{proof}
$(1)\rightarrow(2)$. 
Every unary relation that can be inferred by the 
arc-consistency procedure is definable by a primitive positive formula and hence is an element of $\{O_1,\dots,O_n\}$. For every variable $u$ of $S$, let $T^u$
be the subset of $\{O_1,\dots,O_n\}$ containing all those unary IDBs $O_i$, $1\leq i\leq n$,
such that $O_i(u)$ is derived by $\Pi$. By the structure
of the rules of the arc-consistency algorithm, $T^u$ is closed under intersection. 
Define $h$ to be the mapping from $D_S$ to $\{O_1,\dots,O_n\}$ that maps
every variable $u$ to the minimum element of $T^u$ (with respect to set inclusion), which will be denoted by $\cap T^u$. We shall show that $h$
is a homomorphism from $S$ to the definable subset structure of $\Gamma$. Let $R\in \tau$, and let $(u_1,\dots,u_k)$ be a tuple of $R^S$. Then 
$(\cap T^{u_1},\dots,\cap T^{u_k})$ is the image of this tuple under $h$. Fix any $j \in \{1,\dots,k\}$, and
let $O$ be the set containing all those $v_j$ such that 
there 
are vertices $v_1, \dots, v_{j-1}, v_{j+1}, \dots, v_k$ from $\cap T^{u_1}, \dots, 
 \cap T^{u_{j-1}}, \cap T^{u_{j+1}}, \dots, \cap T^{u_k}$, respectively,
such that $R$ holds on $v_1, \dots, v_k$ in $\Gamma$.
Then $O$ is primitive positive definable in $\Gamma$, and $\Pi$ contains the rule
$$O(x_j) \; {{:}-} \; R(x_1,\dots,x_k), \cap T^{u_1}(x_1),\dots,\cap T^{u_{j-1}}(x_{j-1}),
\cap T^{u_{j+1}}(x_{j+1}),\dots, \cap T^{u_k}(x_k)$$
which allows to derive $O(u_j)$. As $\cap T^{u_j}\subseteq O$ we conclude
that $(\cap T^{u_1},\dots,\cap T^{u_k})$ belongs to the relation $R$ in the definable subset structure.

$(2)\rightarrow(1)$. Let $h$ be a homomorphism from $S$ to the definable subset structure of $\Gamma$. 
It is easy to prove by induction on the evaluation of $\Pi$ on $S$ that $h(u)\subseteq R$ for every $R(u)$ derived by $\Pi$. Hence, $\false$ cannot be derived by $\Pi$ on $S$.
\qed \end{proof}


\begin{theorem}
Let $\Gamma$ be an $\omega$-categorical structure with finite relational signature. 
Then the arc-consistency procedure correctly
decides $\Csp(\Gamma)$ if and only if the definable subset structure is homomorphic to $\Gamma$.
\end{theorem}
\begin{proof}
Since the definable subset structure homomorphically
maps to itself, the claim proven above shows that the
arc-consistency procedure does not derive \false on the definable subset structure.
Hence, if the arc-consistency procedure solves
$\Csp(\Gamma)$ then the definable subset structure homomorphically
maps to $\Gamma$. 

Conversely, suppose that there is a homomorphism $h$
from the definable subset structure to $\Gamma$. To
show that $\Pi$ solves
$\Csp(\Gamma)$, it suffices to show that an instance $S$
where $\Pi$ does not derive \false 
homomorphically maps to $\Gamma$. By the claim proven
above there is a homomorphism $g$ from $S$ to the definable subset structure of $\Gamma$. Composing $g$ and $h$ yields the desired homomorphism
from $S$ to $\Gamma$.
\qed \end{proof}

\begin{theorem}\label{thm:ac}
Let $\Gamma$ be an $\omega$-categorical
structure with finite relational signature. 
If $\Csp(\Gamma)$ is 
solved by the arc-consistency algorithm,
then $\Gamma$ is homomorphically equivalent to a finite structure.
\end{theorem}
\begin{proof}
By Lemma~\ref{lem:infinst}
it suffices to show that arbitrary finite substructures $S$ of
$\Gamma$ homomorphically map to the (finite!) definable subset structure. Since substructures of $\Gamma$ are satisfiable instances of $\Csp(\Gamma)$, $\Pi$ does not derive \false on such a structure $S$.
So by the claim above, $S$ is homomorphic to the definable subset structure of $\Gamma$.
\qed \end{proof}

\section{Bounded strict width}
\label{sect:swk}
The notion of strict width was introduced for finite domain
constraint satisfaction problems by Feder and Vardi~\cite{FederVardi},
and was defined in terms of the canonical Datalog program.
In the terminology of the constraint satisfaction literature in Artificial Intelligence, strict width $l$ is equivalent to 
\emph{`strong $l$-consistency implies global consistency'}.
Based on our generalization of the concept of canonical Datalog programs, we study the analogously defined concept of strict
width $l$ for $\omega$-categorical structures.

The notion of strict width is defined as follows. Recall that the canonical $(l,k)$-Datalog program $\Pi$ for $\Csp(\Gamma)$
receives as input an instance $S$ of $\Csp(\Gamma)$ and returns an expansion $S'$ of $S$ over $\tau'$ where
$\tau'$ is the vocabulary that contains $\tau$ as well as a predicate for every IDB of $\Pi$. The
structure $S'$ can be seen as an instance of $\Csp(\Gamma')$ where $\Gamma'$ is the expansion of $\Gamma$ by all at most $l$-ary primitive positive definable relations. The instance $S'$
is called \emph{globally consistent}, if 
every partial homomorphism, i.e, every homomorphism from an induced substructure of $S'$ to $\Gamma$, can be extended to a homomorphism from $S$ to $\Gamma$.
If for some $k \geq l+1 \geq 3$ all instances of $\Csp(\Gamma')$ 
that are computed by the canonical $(l,k)$-program are globally consistent, we say that $\Gamma$ has \emph{strict width $l$}.
Note that strict width $l$ implies width $l$, and hence
$\Csp(\Gamma)$ can be solved in polynomial time when
$\Gamma$ has bounded strict width.

Also note that if $\Pi$ derives \false\ on input $S$,
then $S'$ does not have any partial homomorphisms to $\Gamma'$, and hence $S'$ is in this case by definition globally consistent. If the reader feels uneasy about calling
unsatisfiable instances globally consistent, one
might also define global consistence only for satisfiable
instances; for strict width $l$ we then require that the instances computed by the canonical $(l,k)$-program that do not contain the predicate $\false$
 are globally consistent.
These two definitions are clearly equivalent. 
With our definition we follow what is standard in the literature.

In this section we present an universal-algebraic characterization of
strict width $l$ for $\omega$-categorical templates $\Gamma$. 
The algebraic approach rests on the
notion of \emph{polymorphisms}.
Let $\Gamma$ be a relational structure with signature $\tau$.
A \emph{polymorphism} is a homomorphism from $\Gamma^l$ to $\Gamma$,
for some $l$, where $\Gamma^l$ is a relational $\tau$-structure
defined as follows.
The vertices of $\Gamma^l$ are $l$-tuples over elements from $V_\Gamma$,
and $k$ such $l$-tuples $(v_1^i, \dots, v_l^i)$, $1 \leq i \leq k$,
are joined by a $k$-ary relation $R$ from $\tau$
if $(v_j^1, \dots, v_j^k)$ is in $R^\Gamma$, for all $1 \leq j \leq l$.


We say that an operation $f$ is a \emph{near-unanimity operation} 
(short, \emph{nu-operation}) if it satisfies the
identities 
$f(x, \dots, x, y, x, \dots, x) = x$, i.e.,
in the case that the arguments have the same value $x$ except at most one argument, the operation has the value $x$.
We say that $f$ is a \emph{near-unanimity operation on $A$} if it satisfies the
identities 
$f(x, \dots, x, y, x, \dots, x) = x$ for all $x,y \in A$.


Feder and Vardi~\cite{FederVardi} proved that a finite template 
$\Gamma$ has an $(l+1)$-ary near-unanimity operation 
(in this case, they say that $\Gamma$ has 
the \emph{$(l+1)$-mapping property}) if and only if
$\Csp(\Gamma)$ has strict width $l$. Another proof
of this theorem was given in~\cite{CCC}.
It is stated there that the proof extends to
arbitrary infinite templates, if we want to characterize
bounded strict width on instances of the constraint
satisfaction problem that might be infinite.
However, we would like to describe the complexity of constraint
satisfaction problems with finite instances.

In fact, there are structures that do not
have a nu-operation, but where $\Gamma$ has bounded strict width.
One example
of such a structure is the universal triangle-free graph $\ntriangleleft$. 
A theorem by Larose and Tardif
shows that every finite or infinite graph with a nu-operation
is bipartite~\cite{Tardif}. 
Since the universal triangle-free graph 
contains all cycles of length larger than three,
it therefore cannot have a nu-operation. 
However, the universal triangle-free graph has strict width $2$. Indeed,
for any instance $S$ accepted by the canonical
$(2,3)$-Datalog program, every partial mapping from $S$ to $\ntriangleleft$ 
satisfying all the facts
derived by the program (and in particular not containing any triangle)
can be extended to a complete homomorphisms from $S$ to $\ntriangleleft$ -- 
this follows from the extension properties of the template.

Theorem~\ref{thm:swk} characterizes strict width $l$, $l\geq 2$,
for constraint satisfaction with $\omega$-categorical templates.
We first need an intermediate result.

\begin{lemma}\label{lem:eq}
Let $\Gamma$ be a $\tau$-structure such that $\Csp(\Gamma)$ has
strict width $l$ and let $\tau_{\equiv}$ be the superset of
$\tau$ in which we add a new binary relation symbol $\equiv$.
Let $\Gamma_{\equiv}$ be the $\tau_{\equiv}$-expansion of $\Gamma$ in which 
$\equiv$ is interpreted by the usual equality relation $\{(x,x) \; | \; x \in D_\Gamma\}$. Then $\Csp(\Gamma_\equiv)$ has also strict width $l$.
\end{lemma}
\begin{proof}
Let $\Pi_{\equiv}$ be the canonical $(l,k)$-program for $\Csp(\Gamma_{\equiv})$,
and $S$ be an instance of $\Csp(\Gamma_{\equiv})$. 
Let $S'$ be the structure computed
by $\Pi_{\equiv}$ on $S$. 
Let $E$ be the smallest equivalence relation on the universe of $S$ that contains ${\equiv}^S$. 
Let $S/E$ be the $\tau$-reduct of $S$ obtained by
factoring $S$ by the equivalence relation $E$.
More precisely, the universe of $S/E$ are the equivalence classes of $R$, $\{E_a\ |\ a\in D_S\}$, where $E_a$ denotes the $E$-class of $a$, and for every $R \in \tau$, say $r$-ary,
$R^{S}=\{(E_{a_1},\dots,E_{a_r})\ |\ (a_1,\dots,a_r)\in R^{S}\}$. We now consider $S/E$ as an instance of $\Csp(\Gamma)$. Let $\Pi$ be the canonical $(l,k)$-program of $\Gamma$.  It is easy to prove by induction on the evaluation of $\Pi$ on $S/E$ that if $R$ is an IDB, say $r$-ary, and $R(E_{a_1},\dots,E_{a_r})$ is derived by $\Pi$ on $S/E$, then $R(a_1,\dots,a_r)$ is derived by $\Pi_{\equiv}$ on $S$.
We have to show that $S'$ is globally consistent.
So suppose that there is a partial homomorphism $h$ from $S'$ to $\Gamma_{\equiv}$. 
Since $l\geq 2$ and $k\geq 3$, $\Pi_{\equiv}$ will
be able to derive that all elements in the same $E$-class have to get the same value and hence, if $h$ is a partial homomorphism then this implies that for all elements $a$, $b$ in the domain of $h$ that are $E$-related, $h(a)=h(b)$. Define $h/E$ to be the partial mapping that maps every $E_a$ with $a$ in the domain of $h$ to $h(a)$. By the definition of $S$ and analysis on the predicates derived by $\Pi$ on $S$ carried out above we know that $h_E$ is a partial homomorphism from $S/E$ to $\Gamma$. Hence 
 $h$ can be extended to a full homomorphism $h'$ from $S/E$ to $\Gamma$.
Finally, the mapping $h'$ defined to be $h'(a)=(h/E)(E_a)$
is a homomorphism from $S$ to $\Gamma$ and hence also from $S'$ to $\Gamma'$. \qed \end{proof}

One of the key properties of structures $\Gamma$
with near unanimity polymorphisms is the following. 

\begin{lemma}\label{lem:extension}
Let $\Gamma$ be a relational $\omega$-categorical structure with maximal arity $k$ and an $(l+1)$-ary polymorphism $f$ 
for every finite subset $A$ such that $f$ is a nuf on $A$. 
Let $\Gamma'$ be the expansion of $\Gamma$ by all $l$-ary primitive positive definable relations, and let $S'$ be an instance of $\Csp(\Gamma')$
computed by the canonical $(l,k)$-Datalog program
for $\Gamma$. Then every partial homomorphism
$h$ from $S'$ to $\Gamma'$ has the property that for every fact $R(u_1,\dots,u_r)$ in $S'$ there
exists a tuple $(d_1,\dots,d_r) \in R^{\Gamma'}$
such that $h(u_i)=d_i$ for all $u_i$ where $h$ is defined. 
\end{lemma}
\begin{proof}
Let $i_1,\dots,i_s$ be a list of the indices $i \in \{1,\dots,r\}$ such that
$u_i \in D_S$, and let $j_1,\dots,j_t$ be a list
of the other indices in $\{1,\dots,r\}$ 
(so we have $s+t=r$). We prove the statement by induction on $s$.  For $s \leq l$, 
let $R'$ be the IDB associated to the 
$\exists u_{j_1},\dots,u_{j_t}. \, R(u_1,\dots,u_r)$
with free variables $u_{i_1},\dots,u_{i_s}$. 
Since $R'(u_{i_1},\dots,u_{i_s}) \; \dlg \; R(u_1,\dots,u_r)$
is a rule in $\Pi$, we have $(h(u_{i_1}),\dots,h(u_{i_s})) \in {R'}^{\Gamma'}$. Then the witnesses for the existentially
quantified variables $u_{j_1},\dots,u_{j_t}$  in $\Gamma'$ along with
$h(u_{i_1}),\dots,h(u_{i_s})$ determine 
the tuple $(d_1,\dots,d_r) \in R^{\Gamma'}$ with the desired property.

For $s \geq l+1$, consider for all
$j \in \{i_1,\dots,i_{l+1}\}$ the tuple $b^j=(b^j_1,\dots,b^j_r) \in R^{\Gamma'}$ given inductively for the restriction of $h$ to
$D_S \setminus \{u_j\}$. Let $g$ be an $l+1$-ary polymorphism which is a nuf on the 
set containing all elements in all tuples $b^j$. Then the tuple
$(g(b_1^1,\dots,b_1^{l+1}),\dots,g(b_r^1,\dots,b_r^{l+1}))$
has the desired properties.
\qed \end{proof}

The proof of the following theorem
is based on ideas from~\cite{FederVardi}
and~\cite{CCC}. 

\begin{theorem}\label{thm:swk}
Let $\Gamma$ be an $\omega$-categorical structure 
with relational signature $\tau$ of bounded maximal arity. 
Then the following are equivalent, for $l\geq 2$:
\begin{enumerate}
\item $\Csp(\Gamma)$ has strict width $l$.
\item For every finite subset $A$ of $\Gamma$ there is an $(l+1)$-ary 
polymorphism of $\Gamma$ that is a nuf on $A$.
\end{enumerate}
\end{theorem}

\begin{proof}
We first show that (1) implies (2).

We assume that $\Csp(\Gamma)$ has
strict width $l$, and prove that for every finite subset $A$ of
$\Gamma$ there is a 
polymorphism of $\Gamma$ that is an $(l+1)$-ary nuf on $A$.
Let $\tau^A$ be the superset of $\tau$ that additionally 
contains a unary relation symbol $R_a$ for each element $a$ of $A$. Let $\Gamma^A$ be the $\tau^A$-expansion of $\Gamma$ 
in which $R_a$ is interpreted by the singleton relation $\{a\}$.
Consider the set $B$ of tuples
$(a_0, \dots, a_l)$ in  $A^{l+1}$ 
that have identical entries 
$a_i=a$ except possibly at one exceptional position. 
Let $\Delta$ be the $\tau^A$-expansion of 
$\Gamma^{l+1}$ where $R_a$ denotes
the set of all tuples $(a,\dots,a,b,a,\dots,a)$ in
$B$ where at most one entry is not $a$.
Every homomorphism from $\Delta$ to $\Gamma^A$ is 
by construction a 
polymorphism of $\Gamma$ that
is  a nuf on $A$. Lemma~\ref{lem:infinst} shows that if 
every finite substructure 
$S^A$ of $\Delta$ homomorphically maps to $\Gamma^A$,
then $\Delta$ homomorphically maps to $\Gamma^A$ as well.

Let $S^A$ be any finite substructure of $\Delta$,
and let $S$ be the $\tau$-reduct of $S^A$, which we
see as an instance of $\Csp(\Gamma)$. We show
that there exists a homomorphism $h$ from $S$ to $\Gamma$ that sends every tuple of the form $(a,\dots,a,b,a,\dots,a)$ in $B \cap D_S$ to $a$. Hence, $h$ is also 
a homomorphism from $S^A$ to $\Gamma^A$.

Let $\tau_{\equiv}$ be the superset of $\tau$ that additionally contains a
new binary predicate $\equiv$, 
and let $\Gamma_{\equiv}$ be the
expansion of $\Gamma$ in which $\equiv$ is interpreted by
the equality relation. 
Let $T$ be the $(\tau_{\equiv})$-structure with domain $D_S \times \{0,1\}$ where an $r$-ary predicate $P \in \tau$ denotes 
$$P^T \; := \; \{((a_1,0),\dots,(a_r,0)) \; | \; (a_1,\dots,a_r) \in P^S\} \; .$$
Furthermore,
$$\equiv^T \; := \; \{((a,0),(a,1))\ |\ a\in S\}.$$

By Lemma~\ref{lem:eq}, $\Gamma_{\equiv}$
has strict width $l$. Let $\Gamma'_\equiv$ be the expansion of $\Gamma_{\equiv}$
by all at most $l$-ary primitive positive definable relations,
and let $k$ be such that all
instances of $\Csp(\Gamma'_{\equiv})$ computed by the canonical $(l,k)$-program 
$\Pi_\equiv$ are globally consistent. 
Let $T'$ be the instance of $\Csp(\Gamma_\equiv')$ 
computed by $\Pi_\equiv$ on $T$. Now consider the
partial assignment $g$ defined on $(B\cap D_S)\times\{1\}$ that sends
every tuple of the form $((a,1),\dots,(a,1),(b,1),(a,1),\dots,(a,1))$ to $a$. We shall see that $g$ is a partial homomorphism from $T'$ to $\Gamma_{\equiv}'$.
Indeed, let $(\overline{a}_1,\dots,\overline{a}_r)\in R^{T'}$ be any tuple 
entirely contained in the domain of $g$. For every $j \in \{1,\dots,r\}$, the tuple $\overline{a}_j$
is of the from $((a_j,1),\dots,(a_j,1),(b_j,1),(a_j,1),\dots,(a_j,1))$. 
This tuple has necessarily
been placed there by the Datalog program, 
and hence $R$ is an IDB and has cardinality at
most $l$. The pigeon-hole principle guarantees that there exists an index $i \in \{1,\dots,l+1\}$ such that for every $1\leq j\leq r$ the
$i$-th entry of ${\overline{a}_j}$ is precisely $(a_j,1)$. Since the $i$-th projection
is a homomorphism from $S$ to $\Gamma$, it cannot violate any fact derived
by the canonical $(l,k)$-Datalog program and hence $(a_1,\dots,a_l)\in R^{\Gamma_{\equiv}'}$.
Since $T'$ is globally consistent this implies that $g$ can be extended to a full homomomorphism
$g'$ from $T'$ to $\Gamma_{\equiv}'$.
Finally we obtain the desired homomorphism $h \colon S\rightarrow D_{\Gamma}$ as 
$h(a_1,\dots,a_l) := g'((a_1,0),\dots,(a_l,0))$.

Next we show that (2) implies (1).
Let $k$ be larger than
the maximal arity of the relations in $\tau$, and at least $l+1$. 
Let $\Pi$ be the canonical $(l,k)$-program for $\Gamma$,
let $\Gamma'$ be the expansion of $\Gamma$ by all at most $l$-ary
primitive positive definable relations, and 
let $S'$ be the instance of $\Csp(\Gamma')$ computed by $\Pi$ on $S$. 
We shall prove 
that every partial homomorphism
with domain $\{v_1,\dots,v_i\}$, 
for $i < |S'|$, has an
extension to any other element $v$ of $S'$
such that the extension is still a partial homomorphism from $S'$ to $\Gamma'$.
We prove this by induction on the size $i$ of the
domain of $s$.

For the case that $i\leq l$,
let $\Psi$ be the set of all atomic formulas
of the form $R(\bar u)$ that hold in $S'$ and
where all entries of $\bar u$ are from
$\{v_1,\dots,v_i,v\}$, and let $R'$ be the IDB
associated to the primitive positive formula 
$\exists v \bigwedge \Psi$ 
with free variables $v_1,\dots,v_i$.
Since each formula in $\Psi$ is derived by $\Pi$ on 
$S$, the predicate $R'(v_1,\dots,v_i)$ is also derived
by $\Pi$ on $S$.
Since $h$ preserves $R'$, we have that 
$(h(v_1),\dots,h(v_i))$ satisfies $\exists v \bigwedge \Psi$; hence, 
there exists an extension of $h$ to $v$ 
such that the extension is a partial homomorphism 
from $S'$ to $\Gamma'$.

For the induction step where $i \geq l+1$, 
select elements $w_1,\dots,w_{l+1}$ 
in $\{v_1,\dots,v_i\}$, and
let $h_j$ be the restriction of $h$ where $w_j$ is undefined, for $j \in \{1,\dots,l+1\}$. 
By induction, $h_j$ can be extended to a homomorphism $h'_j$ from the structure induced
by $\{v_1,\dots,v_i,v\} \setminus \{w_j\}$ in $S'$ to $\Gamma'$. For each $(u_1,\dots,u_r) \in R^{S'}$,
Lemma~\ref{lem:extension} asserts the existence
of a tuple $(b^j_1,\dots,b^j_r) \in R^{\Gamma'}$ 
such that $h'_j(u_i)=b^j_i$ for all $u_i$ where $h$ is defined. 
Let $A$ be the finite set that contains all those elements $b^j_i$ of $\Gamma'$, for all tuples $(u_1,\dots,u_r)$ in all relations $R$ of $S'$.
Let $g$ be an $(l+1)$-ary polymorphism of $\Gamma'$ that is 
a nuf on $A$ (observe that $\Gamma$ and $\Gamma'$ have the same polymorphisms). 
Define $b$ to be
$g(h'_1(v),\dots,h'_{l+1}(v))$. We claim that the
extension $h'$ of $h$ mapping $v$ to $b$ is a homomorphism from
the substructure induced by $\{v_1,\dots,v_i,v\}$ in $S'$
to $\Gamma'$. 

Let $(u_1,\dots,u_r) \in R^{S'}$ be arbitrary;
we want to show that $(h'(u_1),\dots,h'(u_r)) \in R^{\Gamma'}$. Recall that $(b^j_1,\dots,b^j_r) \in R^{\Gamma'}$ 
is such that $h'_j(u_i)=b^j_i$ for all $u_i$ where $h$ is defined. Then the tuple $(g'(b_1^1,\dots,b_1^{l+1}),
\dots,g'(b_r^1,\dots,b_r^{l+1}))$ is from $R^{\Gamma'}$. 
Moreover, we claim that $g'(b_s^1,\dots,b_s^{l+1}) = h'(u_s)$:
if $u_s \in \{v_1,\dots,v_i\}$, note that for all but at
most one $j$ from $\{1,\dots,l+1\}$ we have that $b_s^j = h'_j(u_s) = h(u_s)$,
and since $g'$ is a nuf on the entries of the tuples $b^j$ we obtain that $g'(b_s^1,\dots,b_s^{l+1}) = h(u_s) = h'(u_s)$. 
Otherwise, if $u_s = v$, then $g'(b_s^1,\dots,b_s^{l+1})
= g'(h_1'(v),\dots,h_{l+1}(v)) = b = h'(v)$ by definition of $h'$. 
We conclude that $(h'(u_1),\dots,h'(u_r)) \in R^{S'}$.
\qed
\end{proof}

Note that in several papers including~\cite{OligoClone,qe} and the conference version that
precedes this one, condition (2) has been stated in a different
but essentially equivalent way using the notion of quasi near-unanimity operation.\footnote{In the conference version of this paper, these
operations were called \emph{weak near-unanimity operations}. However, since another similar but much weaker relaxation of near-unanimity operations
was introduced recently in universal algebra as well, we decide to call our operations \emph{quasi} near-unanimity operations.}

We say that an operation $f$ is a \emph{quasi near-unanimity operation} 
(short, \emph{qnu-operation}), if it satisfies the
identities $f(x, \dots, x, y, x, \dots, x) = f(x, \dots, x)$, i.e.,
in the case that the arguments have the same value $x$ except at one argument position, 
the operation has the value $f(x, \dots, x)$.
In other words, the value $y$ of the exceptional
argument does not influence the value of the operation $f$.
Several well-known temporal and spatial constraint languages 
have polymorphisms that are qnu-operations~\cite{qe}.

For every subset $A$ of $\Gamma$, we say that an operation is
\emph{idempotent on $A$} 
if $f(a, \dots, a)=a$ for all $a \in A$.
Hence, if a qnu-operation $f$ is idempotent on the entire domain, 
then $f$ is a near-unanimity operation. 
If a polymorphism $f$ of $\Gamma$ has the property that for
every finite subset $A$ of $\Gamma$ there is an automorphism $\alpha$ of $\Gamma$ such that $f(x,\dots,x)=\alpha(x)$ for all $x \in A$,
we say that $f$ is \emph{oligopotent}.

\begin{corollary}\label{cor:cond}
Let $\Gamma$ be a $\omega$-categorical structure with finite relational
signature $\tau$, and let $l\geq 2$. Then the following are equivalent:
\begin{enumerate}
\item $\Csp(\Gamma)$ has strict width $l$.
\item For every finite subset $A$ of $\Gamma$ there is an $(l+1)$-ary 
polymorphism of $\Gamma$ that 
is a nuf on $A$.
\item $\Gamma$ has an oligopotent $(l+1)$-ary polymorphism
that is a qnu-operation.
\item Every primitive positive formula
is in $\Gamma$ equivalent to a conjunction
of at most $l$-ary primitive positive formulas.
\end{enumerate}
\end{corollary}
\begin{proof}
The equivalence of (1) and (2) has been shown in Theorem~\ref{thm:swk}, and the equivalence of (2) and (3) follows from a direct application of Lemma~\ref{lem:infinst}. The equivalence of (3) and (4) is shown in~\cite{OligoClone}. 
\qed \end{proof}

Concerning the condition of oligopotency 
in statement (3) of Corollary~\ref{cor:cond}, we want to remark that
for every $\omega$-categorical structure $\Gamma$ there is a template that has the same CSP and where all polymorphisms are oligopotent.
It was shown in~\cite{Cores-journal} that every $\omega$-categorical
structure is homomorphically equivalent to a \emph{model-complete
core} $\Delta$, i.e., $\Delta$ has the property
that for every finite subset $A$ of the domain of $\Delta$
and for every \emph{endomorphism} $e$ of $\Delta$ (an endomorphism is a unary polymorphism) there exists an
automorphism $a$ of $\Delta$ such that $a(x)=e(x)$ for all $x \in A$.
(Moreover, it is also known that $\Delta$ is unique up to isomorphism, and $\omega$-categorical.)

\begin{corollary}
Suppose that $\Delta$ is an $\omega$-categorical model-complete core.
Then $\Delta$ has strict width $l$ if and only if $\Delta$ has
an $(l+1)$-ary qnu-polymorphism.
\end{corollary}

\section{Notational link with the relation algebra perspective}
\label{sect:relation-algebra}
This section does not present any new results; instead, it demonstrates how to translate
our results into the terminology of the literature that uses
\emph{relation algebras} to formalize infinite-domain constraint satisfaction problems,
used in particular in temporal and spatial reasoning. 

\subsection{Proper relation algebras}
In Artificial Intelligence, relation algebras are used as a framework to formalize and study qualitative reasoning problems~\cite{LadkinMaddux,Duentsch,HirschAlgebraicLogic}.
In fact, the so-called \emph{network consistency problem} for a fixed relation algebra
turns out to be (up to the way how we formalize the instances of the problem) 
a CSP for a fixed infinite template $\Gamma$.
Relation algebras are designed to handle binary relations in an 
algebraic way; we follow the presentation in~\cite{HirschAlgebraicLogic}. 

\begin{definition}
A \emph{proper relation algebra} is a domain $D$ together with
a set $\mathcal B$ of binary relations over $D$ such that 
\begin{itemize}
\item $\Id := \{(x,x) \; | \; x \in D\} \in \mathcal B$;
\item If $B_1$ and $B_2$ are from $\mathcal B$, then $B_1 \vee B_2 := B_1 \cup B_2 \in \mathcal B$;
\item $1 := \bigcup_{R \in \mathcal B} R \in \mathcal B$;
\item $0 := \emptyset \in \mathcal B$;
\item If $B \in \mathcal B$, then $-B := 1 \setminus B_1 \in \mathcal B$;
\item If $B \in \mathcal B$, then $B^{\smallsmile} := \{(x,y) \; | \; (y,x) \in B\} \in \mathcal B$;
\item If $B_1$ and $B_2$ are from $\mathcal B$, then $B_1 \circ B_2 \in \mathcal B$; where
$$ B_1 \circ B_2 := \{(x,z) \; | \; \exists y ((x,y) \in B_1 \wedge (y,z) \in B_2)\} \; .$$
\end{itemize}
\end{definition}

We want to point out that in this standard definition of proper relation
algebras it is \emph{not} required that $1$ denotes $D^2$.
However, in most examples that we encounter, 
$1$ indeed denotes $D^2$.
The minimal non-empty elements of $\mathcal B$ with respect to
set-wise inclusion are called the \emph{atoms} of the relation algebra, or also the \emph{basic relations}. 

\begin{example}[The Point Algebra]\label{expl:p-pa}
Let $D={\mathbb Q}$ be the set of rational numbers,
and consider 
$$\mathcal B = \{<,>,=,\leq,\geq,\emptyset,{\mathbb Q}^2\} \; .$$
Those relations form a proper relation algebra (with atoms $<,>,=$) which is one
of the most fundamental relation algebras and known under the name \emph{point algebra}.
\qed \end{example}

When $\mathcal B$ is finite,
every relation in $\mathcal B$ 
can be written as a finite union of basic relations, and we abuse
notation and sometimes write $R = \{B_1,\dots,B_k\}$ when
$B_1, \dots,B_k$ are basic relations, $R \in \mathcal B$, and 
$R = B_1 \cup \dots \cup B_k$. 
Note that composition of basic 
relations determines the composition of all relations in the relation algebra, since $$R_1 \circ R_2 = \bigcup_{B_1 \in R_1, B_2 \in R_2}
B_1 \circ B_2 \;.$$

\subsection{Abstract relation algebras}
An \emph{abstract} relation algebra (Definition~\ref{def:rel-algebra} below) is an algebra with signature
$\Id,0,1,-,^{\smallsmile},\vee,\circ$ that satisfies laws that we expect from
those operators in a proper relation algebra.

\begin{definition}[following~\cite{HirschAlgebraicLogic,Duentsch,LadkinMaddux}]\label{def:rel-algebra} An (abstract) relation algebra $\bf A$ is an algebra
with domain $A$ and signature $\{\vee,-,0,1,\circ,^{\smallsmile},\Id\}$ such that
\begin{itemize}
\item the structure $(A; \vee,\wedge,-,0,1)$ is a Boolean algebra where 
  $\wedge$ is defined by $(x,y) \mapsto -(-x \vee -y)$ from $-$ and $\vee$;
\item $\circ$ is an associative binary operation on $A$;
\item $(a^{\smallsmile})^{\smallsmile} = a$ for all $a \in A$; 
\item $\Id \circ~a = a \circ \Id = a$ for all $a \in A$; 
\item $a \circ (b \vee c) =  a \circ b \vee a \circ c$;
\item $(a \vee b)^{\smallsmile} = a^{\smallsmile} \vee b^{\smallsmile}$;
\item $(-a)^{\smallsmile} = -(a^{\smallsmile})$;
\item $(a \circ b)^{\smallsmile} = b^{\smallsmile} \circ a^{\smallsmile}$;
\item $(a \circ b) \wedge c^{\smallsmile} = 0 \; \Leftrightarrow \; (b \circ c) \wedge a^{\smallsmile} = 0$.
\end{itemize}
\end{definition}
We define $x \leq y$ by $x \wedge y = x$.
A \emph{representation} $(D,i)$ of $\bf A$ consists of a set $D$
and a mapping $i$ from the domain $A$ of $\bf A$ to binary relations
over $D$ such that the image of $i$ induces a proper relation algebra $\bf A'$, and $i$ is an isomorphism with respect to the functions 
$\{\vee,-,0,1,\circ,^{\smallsmile},\Id\}$.
In this case, we also say that $\bf A$ is the \emph{abstract relation algebra of $\bf A'$}. 

There are finite abstract relation algebras that do not have a representation~\cite{LyndonRelationAlgebras}. 
Note that when $(D,i)$ is a representation of $\bf A$, then
$i(a)$ is a basic relation of the induced proper relation algebra
if and only if $a \neq 0$, and for every $b \leq a$ we have $b=a$ or $b=0$; we call $a$ an \emph{atom} of $\bf A$. Using the axioms of relation algebras, it can be shown 
that the composition operator is uniquely determined by the
composition operator on the atoms. Similarly, the inverse of an
element $a \in A$ is the disjunction of the inverses of all the atoms below $a$.


\begin{example}\label{expl:a-pa}
The (abstract) point algebra
is a relation algebra with 8 elements and 3 atoms, denoted by $=$, $<$, and $>$. 
The composition operator of the basic relations of the point algebra is shown in the table of Figure~\ref{fig:point-algebra}.
By the observation we just made, this table determines the full composition table.
The inverse of $<$ is $>$, and $\Id$ denotes $=$ which is its own inverse. 
This fully determines the relation algebra.

\begin{figure}
\begin{center}
\begin{tabular}{|l||l|l|l|}
\hline
$\circ$ & $=$ & $<$ & $>$ \\
\hline \hline
$=$ & $=$ & $<$ & $>$ \\
\hline
$<$ & $<$ & $<$ & $1$ \\
\hline
$>$ & $>$ & $1$ & $>$ \\
\hline
\end{tabular}
\end{center}
\caption{The composition table for the basic relations in the point algebra.}
\label{fig:point-algebra}
\end{figure}

We can obtain a representation with domain $\mathbb Q$ 
from the point algebra (Example~\ref{expl:p-pa}) in the obvious way.
Note that this is not the only representation of the abstract point algebra: another representation
can be obtained by taking $[0,1]$ in place of $\mathbb Q$.
While in any representation the relation for $<$ has to be transitive and dense, it need not be unbounded. \qed
\end{example}

\subsection{The network satisfaction problem}\label{sect:network-sat}
The central computational problem that has been studied for relation algebras 
is the 
\emph{network satisfaction problem}~\cite{LadkinMaddux,Duentsch,HirschAlgebraicLogic}.

\begin{definition}
Let $\bf A$ be a finite relation algebra with domain $A$. 
An \emph{(${\bf A}$-)network}~$N = (V,f)$ consists of a finite set of nodes $V$ and a function $f \colon V \times V \rightarrow A$. 
\end{definition}

Two types of network satisfaction problems have been studied for ${\bf A}$-networks.
The first is the \emph{network satisfaction problem for a (fixed) representation $(D,i)$ of
${\bf A}$}: here, the input is an ${\bf A}$-network $N$, and the question is whether $N$
is \emph{satisfiable with respect to $(D,i)$}, that is, whether there exists a mapping $s \colon V \rightarrow D$ 
such that for all $u,v \in V$ 
$$(s(u),s(v)) \in i(f(u,v)) \; .$$

Another problem that has been studied
is the \emph{(general) network satisfaction problem for ${\bf A}$}.
Again, the input is an ${\bf A}$-network $N$. This time the question is whether there
\emph{exists} a representation $(D,i)$ of ${\bf A}$ such that $N$ is satisfiable with respect to $(D,i)$. It is not hard to show that for every finite relation algebra ${\bf A}$ that has
a representation, there is also a representation $(D,i)$ such that the network satisfaction problem for $(D,i)$ is the same problem as the general network satisfaction problem for ${\bf A}$. So we focus on the network satisfaction problem for fixed representations here.



We now present the link between network satisfaction problems and 
constraint satisfaction problems as defined earlier in this paper. 
Let $\tau_{\bf A}$ be a signature
consisting of binary relation symbols: 
$\tau_{\bf A}$ contains a binary relation symbol $R_a$ 
for each element $a \in A$. 
When $(D,i)$ is a representation of $\tau_{\bf A}$, then
we associate to it
a $\tau_{\bf A}$-structure $\Gamma_{D,i}$ in a natural way:
the domain of the structure is $D$, and the relation symbol $R_a$ 
is interpreted 
by $i(a)$. We sometimes also call the $\tau_{\bf A}$-structure $\Gamma_{D,i}$ 
a \emph{representation} of $\bf A$.

Also to each $\bf A$-network $N = (V,f)$ we can associate a 
$\tau_{\bf A}$-structure $S_N$ in a straightforward way:
the domain of $S_N$ is $V$, and for $u,v \in V$ we have
$(u,v) \in R_a$ if and only if $f(u,v)=a$. 
Conversely, we can associate to each finite $\tau_{\bf A}$-structure $S$
a network $N_S = (V,f)$ as follows. 
The node set $V$ of $N$ is $D_S$, the domain of $S$.
Let $u,v \in V$, and list by $a_1,\dots,a_k$ all those elements $a$ of $A$ such that 
$(u,v) \in R_a$.
Then define $f(u,v)=a$ for $a = (a_1 \wedge a_2 \wedge \dots \wedge a_k)$ (if $k=0$, then $a=0$ by definition). 

The following link between the network satisfaction problem for a fixed representation
$(D,i)$ of ${\bf A}$, and the constraint satisfaction problem for $\Gamma_{D,i}$ 
 is straightforward from the definitions.
 
\begin{proposition}\label{prop:netw-sat-wrt-repr}
Let ${\bf A}$ be a finite relation algebra with representation $(D,i)$. 
Then an ${\bf A}$-network $N$ is satisfiable with respect to $(D,i)$ if and only if
$S_N$ homomorphically maps to $\Gamma_{D,i}$. Moreover, 
a finite $\tau_{\bf A}$-structure $S$ homomorphically maps to $\Gamma_{D,i}$
if and only if $N_S$ is satisfiable with respect to $(D,i)$.
\end{proposition}

\subsection{Datalog and Path-Consistency}
One of the main algorithmic techniques used in the context of network satisfaction problems
is the \emph{path consistency procedure}. 
We will see that -- under the translation of terminology presented in Section~\ref{sect:network-sat} -- the path consistency procedure can be formulated with a Datalog program. 


The path-consistency procedure for $\bf A$ takes as input an ${\bf A}$-network $N$. The execution of the procedure on $N$ only depends on ${\bf A}$ as an \emph{abstract} relation algebra
(and not on particular representations of ${\bf A}$). 

\begin{figure}[h]
\begin{center}
\fbox{
\begin{tabular}{l}
PC$_{\bf A}(N)$ \\
Input: an ${\bf A}$-network $N=(V,f)$. \\
Do \\
\hspace{0.5cm} For all distinct nodes $x,y,z \in V$: \\
\hspace{1cm} Replace $f(x,y)$ by $f(x,y) \wedge (f(x,z) \circ f(z,y))$ \\
\hspace{1cm} If $f(x,y)=0$ then {\bf reject} \\
Loop until no further changes \\
Return $(V,f)$. 
\end{tabular}}
\end{center}
\caption{The path-consistency procedure for ${\bf A}$-networks.}
\label{fig:pc}
\end{figure}

\begin{proposition}\label{prop:link}
Let $\bf A$ be a finite relation algebra.
Then there exists a Datalog program $\Pi$ 
such that for every ${\bf A}$-network $N$, 
the program $\Pi$ derives $\false$ on 
$S_N$ if and only if the path-consistency
procedure for $\bf A$ rejects $N$. 
\end{proposition}
\begin{proof}
The Datalog program $\Pi$ is defined as follows. 
The signature $\tau_{\bf A}$ defined above is the 
set of EDBs; as IDBs, we have a binary relation $S_a$ 
for each $a \in A$, and the distinguished $0$-ary predicate \false. Then $\Pi$ contains for each $a \in A$ the rule
$$ S_a(x) \; \dlg \; R_a(x) \; ,$$
and for all $a,b \in A$ the rules
\begin{align*}
S_{a \circ b}(x,y) \; & \dlg \; S_a(x,z), S_b(z,y)  \\
S_{a \wedge b}(x,y) \; & \dlg \; S_a(x,y), S_b(x,y) \; .
\end{align*}
The verification that $\Pi$ has the required properties is straightforward and left to the reader.
\qed \end{proof}

\subsection{Discussion}
We close this section by discussing the weaknesses of the relation
algebra approach to constraint satisfaction. 
First of all, the class
of problems that can be formulated as a network satisfaction problems 
is \emph{severely} restricted.
The relations that we allow in the input network
are closed under unions; this introduces a sort of restricted disjunction 
that quickly leads to NP-hardness, and indeed the network satisfiability problem is tractable in only a few exceptional cases~\cite{HirschAlgebraicLogic}. 
The typical work-around here is to introduce another parameter,
namely a subset of $B$ of the domain of $\bf A$,
and to study the network satisfaction problem for networks
$N=(V,f)$ where the image of $f$ is contained in $B$.
Note that such an additional parameter is not necessary
for CSPs as treated in this paper.
Also note that the network satisfaction problem is restricted to \emph{binary} relations, whereas
many important CSPs can only be formulated in a natural way with
higher-ary relations. As we have seen in Proposition~\ref{prop:link}, every network satisfaction problem for a fixed representation can be formulated
as $\Csp(\Gamma)$ for an appropriate infinite structure $\Gamma$;
but as the above remarks show, only a very small fraction of CSPs
can be formulated as a network satisfaction problem.
 
Even though only very specific CSPs can be formulated
as the network satisfaction problem for a finite relation algebra $\bf A$, 
there are hardly any additional techniques available for
studying the complexity of network satisfaction problems, since the tools we
have for network satisfaction usually also apply to
constraint satisfaction.
For instance, the main computational technique that has been studied for the network satisfaction problem is local consistency (such as path consistency); however, this technique is also applicable
to infinite-domain CSPs in general. 
As we have seen in this paper, 
local consistency is particularly powerful
for problems of the form $\Csp(\Gamma)$ where $\Gamma$ is $\omega$-categorical. 
When the network satisfiability problem under consideration cannot be formulated as $\Csp(\Gamma)$ for an $\omega$-categorical structure $\Gamma$, then not much is known about
the power of consistency techniques for the network satisfiability problem, either. 

The study of composition of
relations in the context of the network satisfiability problem is usually justified by the fact that a network with constraints over the relation $R \circ S$ 
can be simulated by networks that only
have constraints over the relation $R$ and over the relation $S$.
But the same holds for primitive positive definable relations.
Apart from being more powerful, primitive positive definability
has another advantage in comparison to relational composition in relation algebras:
while the set of relations that can be obtained by composing an intersecting the binary relations from a subset of a relation algebra is intricate and not well-understood,
there is a powerful Galois-theory to study primitive positive definability of relations~\cite{BodirskyNesetrilJLC}. In fact, for many infinite structure $\Gamma$ the question
 whether a given first-order formula has a primitive positive definition over $\Gamma$ is decidable~\cite{BodPinTsa}. 

\bibliographystyle{abbrv}
\bibliography{../../global.bib}

\end{document}